\documentclass[useAMS,usenatbib]{mn2e}
\usepackage{graphicx}
\usepackage{amssymb}
\usepackage{amsmath}
\usepackage{color}
\usepackage{lscape}
\usepackage{afterpage}

\title[H$ \alpha $ kinematics of S$ ^{4} $G spiral galaxies-II. {Data description and non-circular motions}]
  {H$ \alpha $ kinematics of S$ ^{4} $G spiral galaxies - II. {Data description and non-circular motions}}
\author[S. Erroz-Ferrer et al.]
{Santiago Erroz-Ferrer,$^{1,2}$\thanks{Email: serroz@iac.es} Johan H. Knapen,$^{1,2}$ Ryan Leaman,$^{1,2,12}$   
\newauthor Mauricio Cisternas,$^{1,2}$ Joan Font,$^{1,2}$ John E. Beckman,$^{1,2}$    Kartik Sheth,$^{3}$
\newauthor  Juan Carlos Mu\~noz-Mateos,$^{4}$ Sim\'on D\'iaz-Garc\'ia,$^{5,6}$  Albert Bosma,$^{7}$
\newauthor  E. Athanassoula,$^{7}$ Bruce G. Elmegreen,$^{8}$ Luis C. Ho,$^{9,10}$
\newauthor  Taehyun Kim,$^{3,4,9,11}$ Eija Laurikainen,$^{5,6}$ Inma Martinez-Valpuesta,$^{1,2}$ 
\newauthor   Sharon E. Meidt,$^{12}$ and Heikki Salo$^{5}$\\
$^{1}${Instituto de Astrof\'isica de Canarias, V\'ia L\'actea s/n 38205 La Laguna, Spain}\\
$^{2}${Departamento de Astrof\'isica, Universidad de La Laguna, 38206 La Laguna, Spain}\\
$^{3}${National Radio Astronomy Observatory / NAASC, 520 Edgemont Road, Charlottesville, VA 22903, USA}\\
$^{4}${European Southern Observatory, Casilla 19001, Santiago 19, Chile}\\
$^{5}${Astronomy Division, Department of Physical Sciences, FIN-90014 University of Oulu, P.O. Box 3000, Oulu, Finland}\\
$^{6}${Finnish Centre of Astronomy with ESO (FINCA), University of Turku, V\"ais\"al\"antie 20, FI-21500, Piikki\"o, Finland}\\
$^{7}${Aix Marseille Universit\'e CNRS, LAM (Laboratoire d'Astrophysique de Marseille) UMR 7326, 13388, Marseille, France}\\
$^{8}${IBM Research Division, T.J. Watson Research Center, Yorktown Hts., NY 10598, USA}\\
$^{9}${The Observatories of the Carnegie Institution for Science, 813 Santa Barbara Street, Pasadena, CA 91101, USA}\\
$^{10}${Kavli Institute for Astronomy and Astrophysics, Peking University, Beijing 100871, China }\\
$^{11}${Astronomy Program, Department of Physics and Astronomy, Seoul National University, Seoul 151-742, Republic of Korea}\\
$^{12}${Max-Planck-Institut f\"ur Astronomie / K\"onigstuhl 17 D-69117 Heidelberg, Germany}}

\date{Accepted 2015 April 23.  Received 2015 March 27; in original form 2014 May 5}

\pagerange{\pageref{firstpage}--\pageref{lastpage}} \pubyear{2015}

\begin{document}

\label{firstpage}

\maketitle

\begin{abstract}
We present a kinematical study of 29 spiral galaxies included in the \textit{Spitzer} Survey of Stellar Structure in Galaxies, using H$ \alpha $ Fabry-Perot data obtained with the Galaxy H$ \alpha $ Fabry-Perot System instrument at the William Herschel Telescope in La Palma, complemented with images in the \textit{R} band and in H$ \alpha $. The primary goal is to study the evolution and properties of the main structural components of galaxies through the kinematical analysis of the FP data, complemented with studies of morphology, star formation and mass distribution. In this paper we describe how the FP data have been obtained, processed and analysed. We present the resulting moment maps, rotation curves, velocity model maps and residual maps. Images are available in FITS format through the NASA/IPAC Extragalactic Database and the Centre de Donn\'ees Stellaires. With these data products we study the non-circular motions, in particular those found along the bars and spiral arms. {The data indicate} that the amplitude of the non-circular motions created by the bar does not correlate with the bar strength indicators. The amplitude of those non-circular motions in the spiral arms does not correlate with either arm class or star formation rate along the spiral arms. {This implies that the presence and the magnitude of the streaming motions in the arms is a local phenomenon.}
\end{abstract}


\section{Introduction}

Galaxies are the basic building blocks of the Universe and understanding their nature, formation and evolution is crucial to many areas of current astrophysical research. Concretely, nearby galaxies are the vivid result of the evolution of the Universe, they contain the footprints of the evolution processes that have led to the present status of the galaxies. To better understand the evolutionary history of the Universe, it is necessary to study the structure and dynamics of current galaxies. By studying the kinematics of galaxies we can trace the motions of both their baryonic (gas and stars) and dark matter. 

Radial velocity measurements have been traditionally performed with radio observations, mainly using the 21-cm H{\sc i} line. Tracing this neutral hydrogen line lets us study most of the gas content of the galaxy, usually traced out to three or four times beyond the visible disc. \citet{Rots1975}; \citet{Bosma1978}; \citet{vanderHulst1979}; \citet{Bosma1981}; \citet{Gottesman1982} and others demonstrated the power of rotation curves in deriving the total mass distribution of disc galaxies. The drawback with these H{\sc i} radio observations was the poor angular resolution. The highest angular resolution ($\sim6''$) 21-cm H{\sc i} surveys of nearby galaxies to date have been carried out by the H{\sc i} Nearby Galaxy Survey (THINGS; \citealt{Walter2008}), using the Very Large Array (VLA) of the NRAO. CO radio observations have also been used traditionally to trace the molecular gas. Nowadays, CO observations provide high angular resolution (comparable to optical), and also high spectral resolution (from one to several km s$ ^{-1} $), but in very small fields.

Optical (H$ \alpha $, NII) observations yield high arcsecond angular resolution data. H$\alpha$ is often the brightest emission line in the visible wavelength range due to the cosmic abundance of hydrogen. In spiral galaxies, this line traces primarily the ionized gas in H{\sc ii} regions around young massive stars. In the 20th century, most of the optical observations were based on long-slit spectroscopy. Long-slit observations have been traditionally used to deduce the rotation curves of galaxies (e.g., \citealt{Rubin1980}; \citealt{Rubin1982a}; \citealt{Rubin1985}; \citealt{Mathewson1992}; \citealt{Persic1995}; \citealt{Mathewson1996}; \citealt{Courteau1997}; \citealt{Dale1997,Dale1998,Dale1999}).

{However, 3D spectroscopy [IFU, Fabry-Perot (FP), multi long-slit spectroscopy] of galaxies is one of the best methods to obtain detailed information on the kinematics of galaxies. Velocity maps derived using 3D spectroscopy reproduce the complete velocity field, contrary to longslit spectra.} H$\alpha$ observations using FP instruments have been used for some 40 years now (\citealt{Tully1974}; \citealt{Deharveng1975}; \citealt{Dubout1976} or \citealt{deVaucouleurs1980}); and have been {designed to derive velocity fields of nearby galaxies (e.g., \citealt{Atherton1982} or \citealt{Boulesteix1984})}, and to create high signal-to-noise ratio (S/N) rotation curves for many spiral galaxies (e.g., \citealt{Marcelin1985}; \citealt{Bonnarel1988}; \citealt{Pence1990}; \citealt{Corradi1991}; \citealt{Amram1992}; \citealt{Cecil1992}; \citealt{Amram1994}; \citealt{Sicotte1996};  \citealt{Ryder1998}; \citealt{Jimenez-Vicente1999}; \citealt{Knapen2000} and many more).

{\citet{Chemin2006} presented an H$ \alpha $ FP survey of Virgo cluster galaxies. \citet{Daigle2006a} also presented an H$ \alpha $ FP survey complementary to the \textit{Spitzer} Infrared Nearby Galaxies Survey (SINGS; \citealt{sings}).} One of the largest and most important FP surveys in H$ \alpha $ is the Gassendi HAlpha survey of SPirals (GHASP; \citealt{Garrido2002}). The GHASP survey consists of a sample of 203 spiral and irregular galaxies that have been observed with a sampling about 16 km s$ ^{-1} $ in velocity and an average 3 arcsec in angular resolution (see \citealt{Epinat2008}; GHASP VII hereafter, for a complete list of data and resolutions). {H$ \alpha $ FP spectrographs are nowadays being  used for studying the kinematics of several kinds of galaxies, e.g., bulgeless galaxies \citep{Neumayer2011}, starburst galaxies \citep{Blasco-Herrera2013}, or interacting galaxies in compact dwarfs \citep{Torres-Flores2014}.} 

{One of the newest FP spectrographs is the Galaxy H$ \alpha $ Fabry-Perot System (GH$\alpha$FaS). It is a visiting instrument on the William Herschel Telescope (WHT) in La Palma. The GH$\alpha$FaS instrument has been operative since 2007 \citep{Fathi2007}, and has been used to study pattern speeds of bars and spiral arms  \citep{Fathi2009}, to analyse the kinematics of planetary nebulae \citep{Santander-Garcia2010}, and the gas flows \citep{Font2011a}, star formation and the kinematics of interacting galaxies (\citealt{Zaragoza2013}, 2014) and starburst galaxies \citep{Blasco-Herrera2013}. It has also been used to study the resonance radii and interlocking resonance patterns in galaxy discs \citep{Font2011b,Font2014}.}

We have designed an observing programme to obtain FP data of 29 spiral galaxies with the GH$\alpha$FaS instrument, as part of the ancillary data of the \textit{Spitzer} Survey of Stellar Structure in Galaxies (S$ ^{4} $G; Sheth et al. 2010). The S$ ^{4} $G survey has obtained 3.6 and 4.5 $ \mu $m images of 2352 nearby galaxies using the Infrared Array Camera (IRAC; \citealt{Fazio2004}). The sample is composed by galaxies that fulfil these requirements: \textit{d} $<$ 40 Mpc, \textit{$m_b$} $<$ 15.5, \textit{$D_{25}$} $>$ 1 arcmin, and includes galaxies selected using values from HyperLeda \citep{Paturel2003}, with a radial velocity v$ _{\rm radio} <$  3000 km s$ ^{-1} $ and galactic latitude $\vert b \vert >$  30$ \degr $. The cornerstone of the S$ ^{4} $G survey is the quantitative analysis of photometric parameters, enabling a variety of studies on secular evolution, outer disc and halo formation, galaxy morphology, etc. The data have been made public\footnote{(http://irsa.ipac.caltech.edu/data/SPITZER/S4G/)}. The first papers resulting from S$^{4}$G survey have been summarised in \citet{Holwerda2014}.

The combination of the S$ ^{4} $G mid-IR images with these complementary kinematic FP data at high resolution will allow us to tackle the scientific goals of this study, summarized as followings:

\begin{enumerate}
\item Perform a detailed study of the kinematical interplay between the interstellar medium and regions of star formation, dust, or other activity.
\item Use the kinematics as a probe of the secular evolution, as manifested in the structural components such as bars, spiral arms, rings, lenses, etc.
\item Exploit the high angular resolution of the kinematic data to study the inner parts of rotation curves, and relate the galaxy kinematics to the mass distribution and specific observed properties, as well as probing the coupling between the stellar density and the gravitational potential in the inner parts of the galaxies.
\item Study possible deviations of the rotation curve that are caused by lopsidedness or asymmetries in the outer parts of the disc, which would help to pin down the interplay between dark matter (DM) and stars.
\item Explore the outer disc kinematics, to establish whether discs are generally cold and thus to constrain the halo properties.
\end{enumerate}

This data paper is the second of a series, and presents the complete set of FP and narrow-band imaging data for the 29 galaxies in our kinematical study. In Paper I (\citealt{Erroz-Ferrer2012}), we illustrated the data and methods, and discussed the kind of results that the main survey will provide through a detailed analysis of NGC~864, an archetypal barred spiral galaxy. In Paper III (Erroz-Ferrer et al.; in prep.) we will present the results from a study of the inner part of the rotation curves of the galaxies of the sample; and Paper IV (Leaman et al.; in prep) will study the outermost parts of the discs of the galaxies of the sample and the relationship between their kinematics and DM.

This paper is organized as follows: Section \ref{section2} gives a description of the sample selection, and Section \ref{section3} describes the observations. Section \ref{section4} describes the data reduction and results. The derived rotation curves and non-circular motions maps are presented in Sections \ref{section5} and \ref{section6} respectively. These results are discussed in Section \ref{section7}, and Section \ref{section8} presents our conclusions.


\section{Target selection}
 \label{section2}

The galaxies in our survey satisfy the following requirements: first of all, the declination should be higher than -10$\degr$ so that the altitude in the sky could be enough at the time of the observation in La Palma, and only galaxies with inclinations between 0$\degr$ and 70$\degr$ were selected.

According to the instrument specifications, the galaxy should fit well in the GH$\alpha$FaS FOV of 3.4 $\times$ 3.4 arcmin. Therefore, galaxies with diameters between 2 and 3.4 arcmin were selected. Secondly, the range in velocity of the galaxy should not be higher than the free spectral range (FSR) of the instrument (see Sect. \ref{section3} for details), although we did not have this ancillary information for all the galaxies. Furthermore, the data reduction processes require that the galaxy should have at least three bright point sources in the field, i.e. three stars or two stars and the nucleus.

The observed sample was selected at the time of the observations, looking at the visibility on the sky, but also requiring a spread in morphological type, bar presence and finally, the availability of ancillary data (interferometric CO, H{\sc i}, ultraviolet or \textit{Spitzer} mid-IR). 

\begin{figure}
\begin{center}
 \includegraphics[width=84mm]{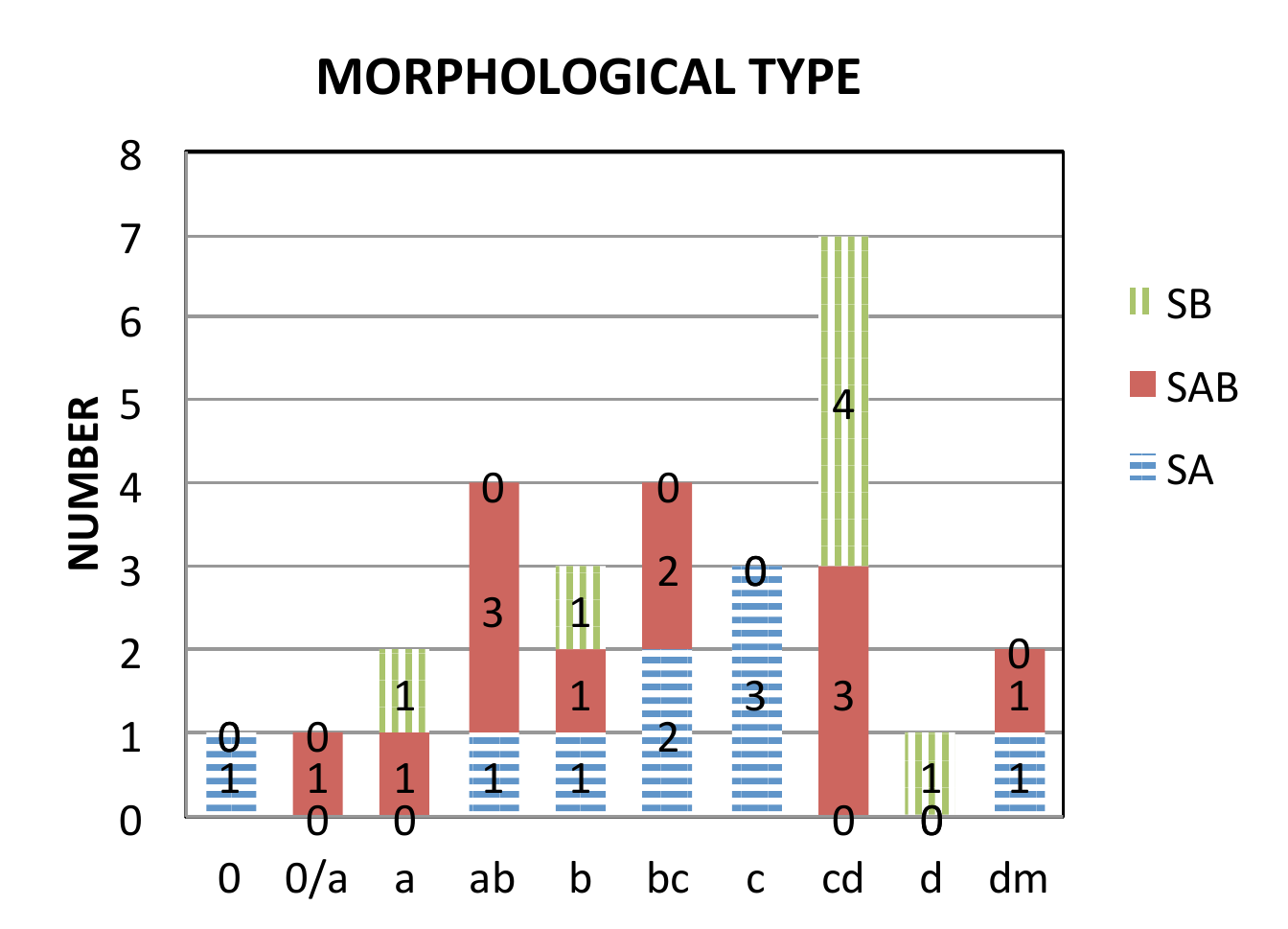}
\caption{Morphological distribution of the galaxies of the sample, organised by morphological type and also by bar presence.}
\label{sample}
\end{center}
\end{figure}

The final sample consists of 29 galaxies. For the morphological classification, we have followed the up-to-date Mid-IR Classifications for S$^{4} $G Galaxies {in the Comprehensive de Vaucouleurs revised Hubble-Sandage System (CVRHS,} \citealt{Buta2015}). The morphological classification also includes bar presence ({9} SA galaxies, 12 SAB galaxies and {7} SB galaxies). In Fig. \ref{sample}, we present the morphological classification as a histogram. The general properties of the galaxies in the sample are presented in Table \ref{props}. As noted in the footnote of Table \ref{props}, the galaxy NGC~7241 is nearly edge-on. The galaxy has an unusual interacting companion in the line of sight, its photometric inclination is biased much lower and therefore we do not include it in the analysis part of this paper. A detailed analysis of it is presented separately in Leaman et al. (in prep.).

\begin{table*}
\caption{General properties of the galaxies in the sample. Notes. (1)~Updated morphological classifications from {the CVRHS}, where ``double stage" galaxies are allowed (i.e., large-scale S0 or S0/a galaxies with smaller-scale inner spirals) (2)~Adopted values of the distances, calculated after applying the Virgo, GA and Shapley corrections, with \textit{H}$_0$= 73 $\pm$ 5 km s$^{-1}$ Mpc$^{-1}$, from the NASA/IPAC Extragalactic Database (NED). The uncertainties in the distance measurements have been adopted as 20 per cent of the value. (3)~\textit{B} magnitude from The Third Reference Catalogue of Bright Galaxies (RC3; \citealt{RC3}). (4)~Absolute \textit{B} magnitude measured using the distance and m$_{b}$ of columns III and IV. (5)~Absolute 3.6$ \mu $m magnitude measured using the distance of column III and asymptotic magnitude at 3.6 $ \mu $m from the ellipse fitting to the 3.6 $ \mu $m S$ ^{4} $G images (Mu\~noz-Mateos et al. in prep). $^{(\dagger)}$ NGC~7241 is nearly edge on, but due to a line of sight companion galaxy, the photometric inclination is biased much lower and should not be considered representative.}
 \label{props}
\center
\begin{tabular}{c|clc|clc|c|}
\hline
  Galaxy name & mid-IR morphology & d(Mpc) & m$_{\rm B}$ & M$_{\rm B}$ & M$_{3.6}$ \\
   & (1) & (2) & (3) & (4) & (5) \\
\hline
  NGC  428 & SAB(s)dm                                                          & 15.9 $\pm$ 3.2& 11.95 & -19.06 & -19.20 \\
  NGC  691 & (R)SA(rs,rl)ab                                                    & 35.7 $\pm$ 7.1& 12.28 & -20.48 & -21.36 \\
  NGC  864 & SAB(r$\underline{\rm s}$)bc                                       & 20.9 $\pm$ 4.2& 11.62 & -19.98 & -20.44 \\
  NGC  918 & SAB(s)cd                                                          & 20.5 $\pm$ 4.1& 13.07 & -18.49 & -20.24 \\
  NGC 1073 & SB(rs)$\underline{\rm c}$d                                        & 16.1 $\pm$ 3.2& 11.68 & -19.35 & -19.75 \\
  NGC 2500 & SAB(s)c$\underline{\rm d}$                                        & 9.8  $\pm$ 2.0& 12.22 & -17.73 & -18.14 \\
  NGC 2541 & SA(s)$\underline{\rm d}$m                                         & 10.4 $\pm$ 2.1& 12.25 & -17.84 & -17.90 \\
  NGC 2543 & SAB(s,bl)b                                                        & 37.4 $\pm$ 7.5& 12.94 & -19.92 & -20.96 \\
  NGC 2712 & (R$^{\prime}$)SAB(rs,nl)a$\underline{\rm b}$                      & 29.5 $\pm$ 5.9& 12.78 & -19.57 & -20.67 \\
  NGC 2748 & (R$^{\prime}$)SAB($\underline{\rm r}$s)bc                         & 25.1 $\pm$ 5.0& 12.39 & -19.61 & -20.84 \\
  NGC 2805 & (R)SA(s)c pec                                                     & 28.7 $\pm$ 5.7& 11.79 & -20.50 & -20.51 \\
  NGC 3041 & SA(rs)$\underline{\rm b}$c                                        & 23.9 $\pm$ 4.8& 12.30 & -19.59 & -20.52 \\
  NGC 3403 & SA(rs)c:                                                          & 22.8 $\pm$ 4.6& 12.94 & -18.85 & -19.63 \\
  NGC 3423 & SA(s)b$\underline{\rm c}$                                         & 14.3 $\pm$ 2.9& 11.60 & -19.18 & -19.58 \\
  NGC 3504 & (R$_1^{\prime}$)SA$\underline{\rm B}$($\underline{\rm r}$s,nl)a & 27.8 $\pm$ 5.6& 11.62 & -20.60 & -21.65 \\
  NGC 4151 & SAB$_a$(l,nl)0/a                                                      & 20.0 $\pm$ 4.0& 11.36 & -20.15 & -21.40 \\
  NGC 4324 & (L)SA(r)0$^+$                                                           & 13.6 $\pm$ 2.7& 12.50 & -18.17 & -19.51 \\
  NGC 4389 & SB(rs)a[d]                                                         & 13.8 $\pm$ 2.8& 12.55 & -18.15 & -19.11 \\
  NGC 4498 & SB(r$\underline{\rm s}$)d                                         & 14.1 $\pm$ 2.8& 12.77 & -17.98 & -18.51 \\
  NGC 4639 & (R$^{\prime}$)SA$\underline{\rm B}$(rs,bl)ab                      & 13.9 $\pm$ 2.8& 12.19 & -18.53 & -19.39 \\
  NGC 5112 & SB(s)c$\underline{\rm d}$                                         & 20.2 $\pm$ 4.0& 12.63 & -18.90 & -19.49 \\
  NGC 5334 & SB(rs,x$_1$r)cd                                                   & 24.2 $\pm$ 4.8& 12.88 & -19.04 & -20.03 \\
  NGC 5678 & (R$^{\prime}$L)SA($\underline{\rm r}$s)b: pec                     & 33.3 $\pm$ 6.7& 12.02 & -20.59 & -21.87 \\
  NGC 5740 & ($\underline{\rm R}$L)SAB(r)ab                                    & 27.0 $\pm$ 5.4& 12.60 & -19.56 & -20.65 \\
  NGC 5921 & SB($\underline{\rm r}$s)b                                         & 26.2 $\pm$ 5.2& 11.68 & -20.41 & -21.31 \\
  NGC 6070 & SA(r$\underline{\rm s}$,nrl)c                                     & 33.6 $\pm$ 6.7& 12.42 & -20.21 & -21.70 \\
  NGC 6207 & SAB(r$\underline{\rm s}$)c$\underline{\rm d}$                     & 18.5 $\pm$ 3.7& 11.86 & -19.48 & -19.84 \\
  NGC 6412 & SB(rs)cd                                                          & 23.7 $\pm$ 4.7& 12.38 & -19.49 & -20.07 \\
  NGC 7241 & S$\underline{\rm c}$d  sp        / E(d)7                       & 22.4 $\pm$ 4.5& 13.23 & -18.52 & -20.51 \\
  \hline
\end{tabular}
\end{table*}


\section{Observations}
 \label{section3}

The Fabry-Perot observations were carried out using the GH$ \alpha $FaS instrument mounted on the 4.2-m WHT in La Palma. The observations were done during 24 nights between September 2010 and March 2013.

The instrument is a Fabry-Perot interferometer that provides high spectral resolution and seeing-limited angular resolution data, within a 3\farcm4 $\times$ 3\farcm4 FOV. The instrument comprises a focal reducer, a filter wheel, a Fabry-Perot etalon, an image photon-counting system (IPCS), and a calibration lamp (neon source). The etalon used has an interference order of 765 and a FSR of 391.9 kms$^{-1}$ (8.6 \AA). Each observational cycle consists of 48 steps (named channels) through the etalon for 10 seconds each. The number of cycles depends on the galaxy magnitude, but usually amounts to some 3 hours (corresponding to 22 cycles). However, due to weather and instrumentation problems, not all the galaxies were observed for the same length of time (all the information about the observations is collected in Table \ref{obs}). The high spectral resolution mode was used, achieving a velocity {sampling} of $\sim$8 km s$ ^{-1} $ with a pixel scale of 0\farcs2 in 1K$\times$1K pixel images. In Table~\ref{setup}, we summarize the instrumental setup used in the observations.

To perform flux calibration, H$ \alpha $ narrow-band images were observed together with the Fabry-Perot observations, using the Auxiliary port CAMera (ACAM), permanently mounted at a folded-Cassegrain focus of the WHT \citep{Benn2008}. The FOV in imaging mode is 8 arcmin, with a pixel size of 0.25 arcsec. The \textit{R}-band images were taken using a Bessel filter with central wavelength 6500\AA{} and full width half maximum (FWHM) of 1360\AA{}, and with a Sloan 6228/1322 filter (central wavelength/FHWM, in \AA{}). We used four different H$\alpha$ filters  depending on the galaxy's redshift (6577/23, 6589/24, 6613/24 and 6631/17; central wavelength/FHWM, in \AA{}). The exposure times per galaxy were 3$ \times $20 seconds for the \textit{R}-band images and 3$ \times $100 seconds for the H$\alpha$ images. The typical seeing was around 1 arcsec, ranging from 0\farcs6 to 1\farcs6. 

\begin{table*}
\caption{Journal of observations.}
 \label{setup}
\center
\begin{tabular}{c|clc|c}
\hline
  Observations & Telescope & 4.2-m WHT\\
 \hline
 Fabry Perot &Instrument  & GH$ \alpha $FaS\\
  & Spatial sampling & FOV & $202"\times202"$\\
  & &  Pixel scale & 0\farcs2 \\
&  Calibration & Neon comparison light & $\lambda$ {6598.95} \AA \\
& Characteristics @ H$ \alpha $ & Interference order & 765\\
& & FSR & 391.9 km s$ ^{-1} $ (8.66 \AA)\\
& & Finesse  &  20\\
& & Spectral resolution & 18179 \\
& &Instrumental FWHM & 19.6 km s$ ^{-1} $\\
& Spectral sampling & Number of scanning steps & 48 \\
& & Sampling step & 8.2 km s$ ^{-1} $ (0.18 \AA)\\
& Detector & IPCS & \\
         \hline
Imaging &Instrument & ACAM\\
  & Spatial sampling & FOV & 8 arcmin\\
  & &  Pixel scale & 0\farcs25 \\
&  Calibration & Standard stars& \\
         \hline
\end{tabular}
\end{table*}

\begin{table*}
\caption{Log of the observation. Notes: (1) The adopted format for the filters is  \textit{central wavelength/FHWM} (2) The seeing was measured from the ACAM images. To perform the continuum subtraction, the H$ \alpha $ image or the \textit{R}-band image were degraded to the same spatial resolution (the worst seeing of the two images), and this number has been adopted as the final seeing.}
 \label{obs}
\center
\begin{tabular}{c|clc|clc|cl}
\hline
  Galaxy name & Date & FP t$ _{exp} $& FP H$ \alpha $ filter$ ^{(1)} $ & ACAM H$ \alpha $ filter$ ^{(1)} $ & seeing$ ^{(2)} $\\
  &  & (s) & (\AA/\AA) &  (\AA/\AA) &  (")\\
\hline
  NGC  428 & Sept. 2010  & 3840   & 6580/23     & 6589/24 & 1.0 \\
  NGC  691 & Jan. 2012   & 11040  & 6623/23     & 6631/17 & 1.2 \\
  NGC  864 & Sept. 2010  & 13920  & 6600.5/25   & 6589/24 & 0.9 \\
  NGC  918 & Sept. 2010  & 2880   & 6600.5/25   & 6589/24 & 0.9 \\
  NGC 1073 & Sept. 2010  & 20640  & 6580/23     & 6589/24 & 0.9 \\
  NGC 2500 & Jan. 2012   & 10560  & 6580/23     & 6577/23 & 1.0 \\
  NGC 2541 & Jan. 2012   & 10560  & 6580/23     & 6577/23 & 1.2 \\
  NGC 2543 & Jan. 2012   & 12960  & 6623/23     & 6613/24 & 1.3 \\
  NGC 2712 & Feb. 2012   & 10560  & 6600.5/25   & 6613/24 & 1.4 \\
  NGC 2748 & Jan. 2012   & 9600   & 6600.5/25   & 6589/24 & 1.4 \\
  NGC 2805 & Mar. 2013   & 9120   & 6597.5/17.6 & 6613/24 & 1.1 \\
  NGC 3041 & Jan. 2012   & 10080  & 6600.5/25   & 6589/24 & 1.4 \\
  NGC 3403 & Jan. 2012   & 10560  & 6580/23     & 6589/24 & 1.1 \\
  NGC 3423 & Feb. 2012   & 9120   & 6580/23     & 6589/24 & 1.5 \\
  NGC 3504 & Jan. 2012   & 15360  & 6600.5/25   & 6589/24 & 1.4 \\
  NGC 4151 & Feb. 2012   & 13920  & 6580/23     & 6589/24 & 1.2 \\
  NGC 4324 & Mar. 2013   & 8160   & 6597.5/17.6 & 6613/24 & 1.4 \\
  NGC 4389 & Jan. 2012   & 12000  & 6580/23     & 6577/23 & 0.8 \\
  NGC 4498 & Jun. 2011   & 10080  & 6598/18     & 6589/24 & 0.9 \\
  NGC 4639 & Jun. 2011   & 8640   & 6583/15.5   & 6589/24 & 1.0 \\
  NGC 5112 & Jun. 2011   & 9120   & 6583/15.5   & 6589/24 & 0.6 \\
  NGC 5334 & Jun. 2011   & 11520  & 6598/18     & 6589/24 & 1.2 \\
  NGC 5678 & Jan. 2012   & 10560  & 6600.5/25   & 6613/24 & 1.2 \\
  NGC 5740 & Jun. 2011   & 11040  & 6598/18     & 6589/24 & 1.5 \\
  NGC 5921 & Mar. 2013   & 7200   & 6597.5/17.6 & 6589/24 & 1.3 \\
  NGC 6070 & Jun. 2011   & 9600   & 6608/16     & 6613/24 & 1.5 \\
  NGC 6207 & Jun. 2011   & 8640   & 6583/15.5   & 6589/24 & 1.3 \\
  NGC 6412 & Jun. 2011   & 13920  & 6598/18     & 6589/24 & 1.3 \\
  NGC 7241 & Jun. 2011   & 7680   & 6598/18     & 6589/24 & 1.1 \\
       \hline
\end{tabular}
\end{table*}


\section{Data reduction}
 \label{section4}

\subsection{H$ \alpha $ imaging: ACAM}
\label{ACAM}
The ACAM images were reduced initially using {\sc iraf}. First, bias and flat corrections were made. Then, the sky was subtracted before and after the combination of the exposures. The images were astrometrically calibrated using {\sc koords} in {\sc kappa} first, setting a preliminar astrometry using the DSS images as a reference. Afterwards, we used {\sc gaia} in {\sc starlink} to improve the astrometry with the GSC-2 catalogue at ESO. The next step was aligning the \textit{R}-band and H$ \alpha $ images for each galaxy, subtracting the continuum using the procedures outlined in \citet{Knapen2004} and \citet{Sanchez-Gallego2012}. Finally, the continuum-subtracted H$ \alpha $ images were flux calibrated using spectrophotometric standard stars observed at the time of the observations.

The resulting continuum-subtracted image of NGC~2748 is presented in panel f) of Fig. \ref{ngc2748plots}. The images of the other galaxies in the sample are presented in Appendix A.

\subsection{Fabry-Perot: GH$ \alpha $FaS}

The basic custom reduction of GH$ \alpha $FaS data cubes has been explained in previous papers in the literature (e.g., \citealt{Hernandez2008} or \citealt{Blasco2010}), and was introduced in Paper I. {The data of all the galaxies have been reduced} following these steps: de-rotation,  {phase-correction}, combination and wavelength calibration, astrometry correction, spatial smoothing, continuum subtraction, removing sky lines, flux calibration and creation of the moment maps.

\subsubsection{De-rotation, {phase-correction}, combination and wavelength calibration}

There is no suitable de-rotator at the Nasmyth focus of the WHT. Therefore, the first step is to de-rotate the data cubes. To do this, we  followed \citet{Blasco2010}. At least two point sources in all the planes are selected (ideally stars), and selected throughout the cubes. After that, each plane of the subsequent cubes is de-rotated to the same position.  

Due to the nature of the FP data, any 2D spatial transformation (rotation or translation) has to be done also in the third dimension (the spectral dimension), and therefore de-rotation must be done simultaneously with the phase calibration. The wavelength calibration is also performed at this point. Lamp exposures were taken between galaxy observations and are used to calibrate in wavelength (see \citealt{Carignan2008}). The integration of all the cubes from the various cycles is performed after the de-rotation and phase calibration. The result is one 48-channel cube, {phase-corrected} and calibrated in wavelength.

\subsubsection{Astrometry calibration}
 The data are placed on an astrometrically correct spatial grid by comparing the positions of stars in the ACAM H$ \alpha $ images. The astrometric calibration of the ACAM images is explained in Sect. \ref{ACAM}.
 
\subsubsection{Spatial smoothing}

The instrument delivers data cubes with a spatial scale of 0\farcs2/pix, and therefore the spatial resolution is limited by the seeing. If the de-rotation process is not correctly performed, the spatial resolution might change (i.e., the addition of misaligned cubes would result in a blurred cube, the PSF of the point source rises and therefore the seeing increases). Thus, the de-rotation output has been carefully checked in order to minimise changes in the resolution.

We applied a 2D Gaussian smoothing kernel with a FWHM of 3 pixels to the data in order to improve the signal-to-noise ratio (S/N) without degrading the angular resolution too much. To apply this Gaussian kernel, we have used the {\sc IDL} task {\sc filter\_image.pro}.

\subsubsection{Continuum subtraction}

The resulting de-rotated, {phase-corrected} and wavelength-calibrated data cube is then imported into {\sc gipsy} (Groningen Image Processing System; \citealt{vanderHulst1992}). The continuum was subtracted using {\sc conrem} of {\sc gipsy}, which estimates the continuum level from line-free frames. To determine the channels free of line emission, we visually check the images and also the spectra, after additional smoothing.
 
{For NGC~2748, NGC~5678 and NGC~6070, the velocity amplitude is equal or larger than the FSR and there are no channels free of line emission. The H$ \alpha $ line falls into the next interference order and reappears in the first or last channels (what we call ``peak intrusion"). Those channels containing emission from a peak intrusion are copied after or before the first or last channel, where they should really be, increasing the number of channels per cube. Then, we divided the cube spatially in two parts (roughly corresponding to the areas where emission from the receding and approaching halves is seen). The channels free of emission were identified separately for each sub-cube, and the continuum was calculated and subtracted separately in each sub-cube. After the continuum subtraction, the two cutouts have been combined again.}

\subsubsection{Sky line removal}

Stationary secondary peaks sometimes appear in the data cubes, usually located in three channels, peaking in the middle one and with a circular light gradient along the image (more light in the centre and gradually less light as leaving the centre). {These are} OH sky emission lines due to airglow, peaking near the following velocities/wavelengths: 690 km s$^{-1}$/6577.3 \AA ~(present in NGC~2500, NGC~2541, NGC~4389 and NGC~6207), 1080 km s$^{-1}$ / 6586.5 \AA ~(NGC~428, NGC~1073, NGC~3403, NGC~4151, NGC~4639 and NGC~5112) and 1480 km s$^{-1}$/6596.7 \AA ~(NGC~4498, NGC~5334 and NGC~6412). To remove this contribution, we have dealt with the channels as if they were images to correct for flat fielding. With {\sc flat} of {\sc gipsy}, we fitted a polynomial to the background, eliminated the gradient across the affected planes and flattened the background.

\subsubsection{Flux calibration}

We calibrate the Fabry-Perot data by using calibrated narrow-band H$ \alpha $ imaging. As explained in Sect. \ref{section3}, the kinematic data have been observed together with ACAM data. The procedure was explained in Paper I: fluxes from selected H{\sc ii} regions in both ACAM and H$ \alpha $ FP data are compared, and fitted to a linear relationship. As a consequence, the zeroth moment maps (intensity maps, see the following section) are flux-calibrated. {GHASP VII uses a different method to flux-calibrate their FP observations. Instead of carrying out their own observations, they compare the integrated H$ \alpha $ flux of their galaxies to that of the narrow-band observations in \citet{James2004}. Although our method requires more observing time at the WHT, it is more accurate} as both FP and narrow-band imaging observations are carried out under the same atmospheric conditions and airmass. 

\subsubsection{Moment maps}

To create the moment maps we have used the {\sc moments} task of {\sc gipsy}, {which performs an intensity weighted mean of the physical coordinates along the profile{\footnote{https://www.astro.rug.nl/$\sim$gipsy/tsk/moments.dc1}}}. We imposed the condition that the line emission had to be present in at least three adjacent channels and at a level above a certain noise level  $ \sigma $ in the profiles. To do that, we determined $\sigma$ for each galaxy using {\sc stat} in {\sc gipsy}, and we created the sets of moment maps with emission above 5$ \sigma $. Also, we  imposed the condition that secondary peaks are not taken into account (although secondary peaks are assumed not to be present after the previous reduction processes).

Subsequently, from the smoothed and continuum-subtracted cube we computed the moment maps for each galaxy, specifically the moment maps of order zero (intensity map), order one (velocity map along the line of sight) and order two (velocity dispersion maps). {The velocity dispersion maps will be used in Paper III, and they will be corrected for instrumental, thermal and natural line broadening. The natural width has a value of 3 km s$ ^{-1} $ \citep{ODell1988}. The thermal width corresponds to 9.1 km s$ ^{-1} $, assuming a temperature of 10$ ^{4} $ K \citep{Osterbrock2006}. The instrumental width for each galaxy was obtained from the data cube taken with the calibration lamp, following the procedures in \citet{Relaño2005}, and is 8.3~km~s$ ^{-1} $.}

After creating the moment maps, the stars and central regions may leave a residual (i.e., the continuum may not be properly determined and  removed), and therefore the outputs from {\sc moments} need to be masked. Specifically, for the AGN hosts NGC~4151 and NGC~4639, the central emission saturated the detector and in consequence the continuum subtraction could not be removed properly, leaving a residual that had to be masked out. Also, we have masked the central regions in NGC~4324, NGC~5740 and NGC~5921, since the signal there is due to continuum emission (i.e., no line emission is detected).

\begin{figure*}
\begin{center}
 \includegraphics[width=170mm]{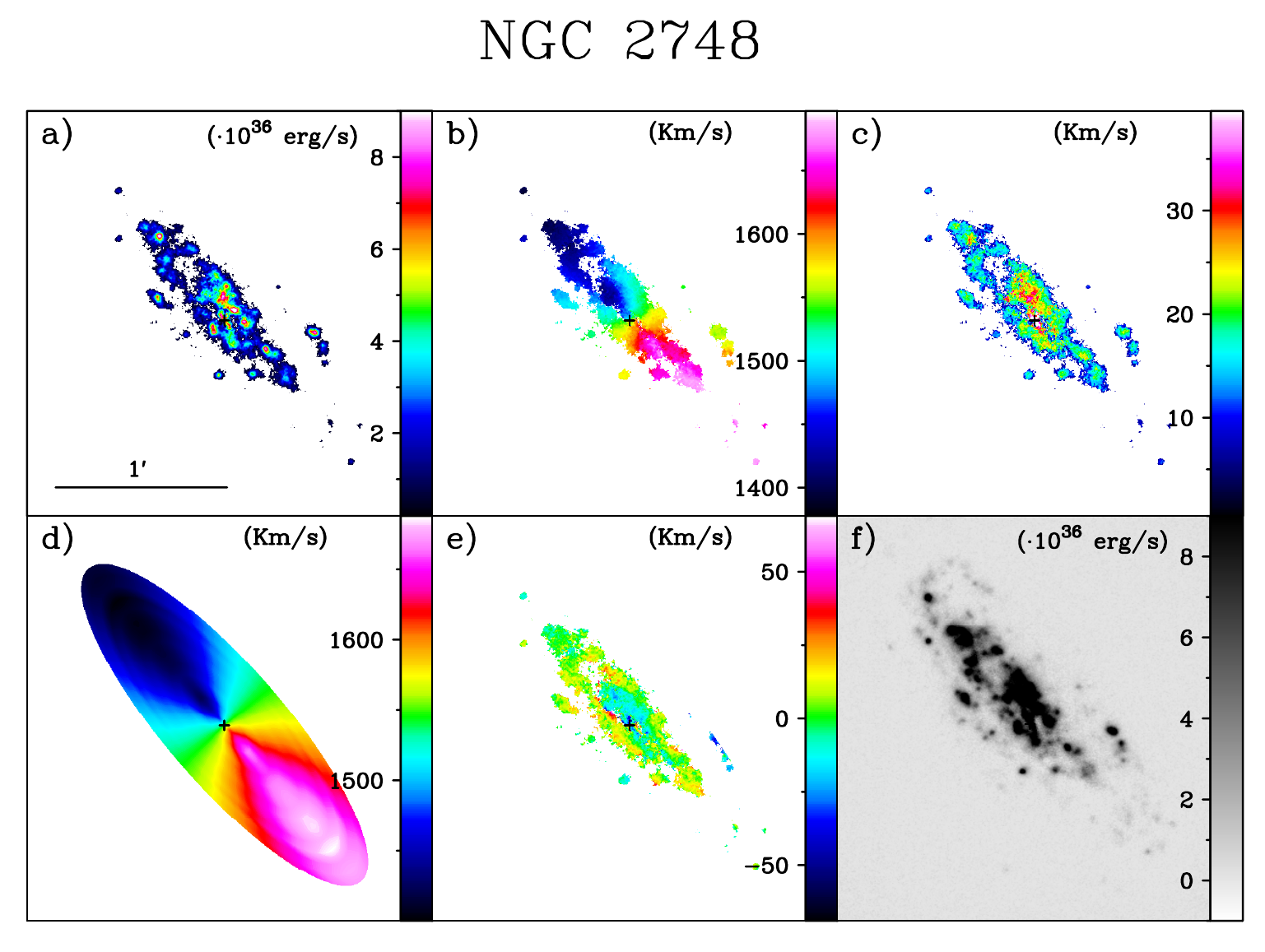}
\caption{Results from the analysis of the Fabry-Perot data cubes of the spiral galaxy NGC~2748.\textit{a)} H$ \alpha $ intensity map. \textit{b)} Velocity map. \textit{c)} Velocity dispersion map. \textit{d)} Velocity model map. \textit{e)} Non-circular motions map. \textit{f)} H$ \alpha $ narrow-band image from ACAM. In the images, North is up and East to the left. All the images have the same scale. Similar panels for all the sample galaxies are presented in App. A.}
\label{ngc2748plots}
\end{center}
\end{figure*}

For each galaxy, we created a six-plots figure, presenting the resulting velocity maps and the outputs from the analysis. In Fig. \ref{ngc2748plots}, we present the resulting moment maps for NGC~2748 in the top images. The figures for the other galaxies in the sample are collected in Appendix A.

\section{Data analysis}

\subsection{Rotation curves}
 \label{section5}

To understand the circular motions within a galaxy, rotation curves have been used widely. Here, we study the rotation of the gas, and in particular, the gas from the H{\sc ii} regions that have been heated by massive, young stars. To extract the rotation curve from the velocity map we have used the {\sc rotcur} task in {\sc gipsy}. This procedure is based on the tilted-ring method described by \citet{Begeman1989}, where each ring can be defined by the parameters: inclination ($i$), position angle (PA), centre ($x_{0}$ and $y_{0}$) and systemic velocity ($v_{\rm sys}$). If we assume that the radial and vertical velocities are negligible, the observed velocity can be expressed as $v_{obs}(R,\theta,i) = v_{sys}+v_{rot}(R,\theta) \cos \theta \sin i$, where $ \theta $ is the azimuthal angle in the plane of the galaxy and depends on  $i$, PA, $x_{0}$ and $y_{0}$.

It is not easy to determine the rotation curve parameters, as deviations from rotational motions can be present inside the galaxies, some of our galaxies are not much inclined, and most of the velocity fields are sparsely sampled. However, we have proceeded to ensure that we are obtaining the true rotation curve while minimising possible errors. In order to determine the starting values (i.e., initial conditions) of $ i $, PA, $x_{0}$, $y_{0}$ and $v_{\rm sys}$, our first step was to search the literature (RC3, \citealt{RC3}; HyperLeda, \citealt{Paturel2003}; GHASP VII; and Mu\~noz-Mateos et al. in prep). We then derived position-velocity diagrams (PV diagrams) along the kinematic major axis of the galaxies in order to see the rotation in the spatial direction (presented as supplementary material in the Appendix C.

Then, we proceeded with {\sc rotcur} as usual: (1) All parameters were left free. (2) To obtain the position of the galaxy centre, the PA, inclination, and systemic velocities were fixed, and the centre values were free. In the majority of the cases, the fits were unsatisfactory due to the patchiness of the data. In the galaxies whose centre is represented by a bright point source (nucleus), we adopted that as the dynamical centre. In the cases where the fits were unsatisfactory and the centre is not presented with a point source, we adopted the value from NED. In any case, the centre positions differ from those given in NED by less than 0\farcs8, less than the angular resolution and less than the accuracy of the astrometry in our data cubes. (3) The centre position, PA, and inclination were fixed in order to fit the systemic velocity.  (4) The PA and inclination were then left free with all other parameters fixed. (5) Finally, the rotation curve was obtained by fixing all parameters except the rotation velocity.

In all the steps, we avoided the sector under 30$\degr$ from the minor axis and a $ \vert cos (\theta)\vert $ weight was applied to the line-of-sight velocities during the fittings. The objective was to minimize the errors caused by the points close to the minor axis, where important projections effects occur and where the circular velocity term is fitted with difficulty $\left[cos (\theta)\rightarrow 0 \right] $. To determine (fix) one parameter, we forced the mean of the values of the rings to satisfy the following conditions, {based on our experience and linked to the resolution of our data}: (i) The number of points in each ring should be greater than a sufficient number, usually 20-50 {(20 is the minimum number of points in a region with the size of our angular resolution, which is $\sim$1 arcsec)}. (ii) The difference between the average of all the values from all the rings and the value in the given ring should not be more than 15 km s$ ^{-1} $ for $v_{\rm sys}$ and 15$\degr$ for the PA and $ i $ {(15 km s$ ^{-1} $ is our velocity resolution, twice the velocity sampling)}. (iii) The uncertainty coming from the free values in the fit (i.e., $v_{\rm sys}$ and the rotational velocity in step 2) should not be more than 10 km s$ ^{-1} $ for the velocities and 10$\degr$ for the PA and $ i $. Usually, these conditions coincided to pick out those cases where the fit was not satisfactory, and the corresponding values were not taken into account in the average. { All the rotation curves have been computed taking into account both approaching and receding sides, and the error bars are the dispersion of the rotation velocities computed for all the pixels in each elliptical ring.}

To check that the fixed values represent the true rotation, we looked at the residual map (see the following section) to search for systematic errors. For example, if the residual map tends to have negative values, it is because the fixed systemic velocity is too large. Other systematic effects that appear in the residual map when one parameter is badly selected are presented in Fig.~{8 of \citet{Warner1973}}.  {Also, we superimposed on the PV diagrams along the major axis the deprojected rotation curve at the corresponding angle, confirming that the derived rotation curves are correct (see Appendix C.}

We have created two types of rotation curves: a high-resolution curve (with a separation of 1" between points) and a low-resolution curve (with a separation of 5"). The high resolution curve traces better the behaviour of the curve and the small-scale features of the galaxy, although it is noisier as there are fewer pixels taken into account in the fit for each radial point.

\begin{figure}
\begin{center}
 \includegraphics[width=84mm]{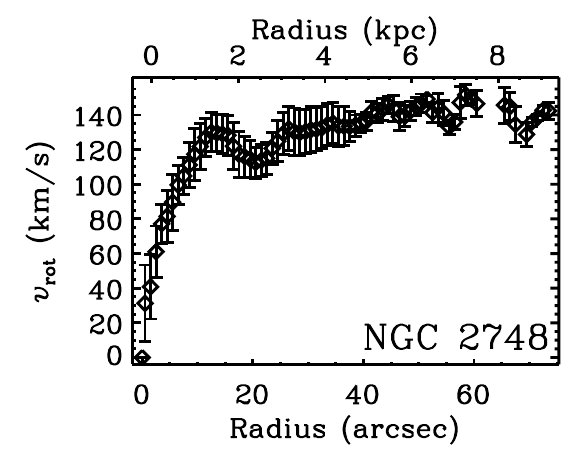}
\caption{High-resolution rotation curve derived from the 5-$ \sigma $ velocity map of NGC~2748. Rotation curves for all the sample galaxies are presented in the Appendix B.}
\label{ngc2748rotcur}
\end{center}
\end{figure}

The parameters ($i$, PA and $v_{\rm sys}$) resulting from the low-resolution fit are presented in Table \ref{rotcuroutput}. For NGC~2748, the high-resolution rotation curve is presented in Fig. \ref{ngc2748rotcur}. The resulting high-resolution rotation curves for all the sample galaxies are presented in the Appendix B.

\begin{table*}
\caption{Results. Column I) Galaxy name. Column II) Systemic velocity from optical observations (HyperLEDA). Column III) Systemic velocity derived from our FP data. Columns IV and VI) Disc inclination and PA obtained from the 25.5 mag/arcsec$ ^{2} $ isophote from the ellipse fitting to the S$ ^{4} $G 3.6 $ \mu $m image (Mu\~noz-Mateos et al. in prep). Columns V and VII) The same as columns IV and VI but derived from our FP data. {Column VIII) Maximum velocity derived from the rotation curves. Column IX) Quality flag for the maximum velocity: reached in our data (1), probably reached (2), probably not reached (3), not reached (4). }\textit{Notes}: (1) PA of the major axis is defined as the angle, taken in anti-clockwise direction between the north direction on the sky and the kinematical major axis of the galaxy, defined from 0\degr to 180\degr. (2) For NGC~1073, 180$\degr$ have been added to PA$_{\rm 3.6 \mu m}$ for a direct comparison with PA$_{\rm H\alpha }$. (3) The AGN contribution has been excluded from the measurements.}
 \label{rotcuroutput}
\centering
\begin{tabular}[angle=90]{c|clclc|clcclclc|clc|clc|c|}
\hline
 Galaxy name& $v_{\rm sys,Leda}$ & $v_{\rm sys,H\alpha }$ & $i_{\rm 3.6 \mu m}$& $i_{\rm H\alpha }$ & PA$_{\rm 3.6 \mu m}^{(1)}$& PA$_{\rm H\alpha }^{(1)}$ & $v_{\rm max}$ & $v_{\rm max}$ \\ 
&  (km s$ ^{-1} $) &  (km s$ ^{-1} $) & $ (\degr)$& ($\degr$) & $(\degr)$& ($\degr$) & (km s$ ^{-1} $) & flag\\
\hline
  NGC  428&1162.4&1150 $\pm$ 6&40.3&45 $\pm$ 9 & 113.9       & 120 $\pm$ 4& 127 $\pm$ 5 & 2\\ 
  NGC  691&2665.0&2712 $\pm$ 5&41.3&41 $\pm$ 10& 93.7        & 91  $\pm$ 4& 245 $\pm$ 3 & 1\\ 
  NGC  864&1558.9&1550 $\pm$ 4&44.6&43 $\pm$ 5 & 22.0        & 25  $\pm$ 6& 169 $\pm$ 8 & 1\\ 
  NGC  918&1502.4&1507 $\pm$ 3&54.6&57 $\pm$ 4 & 157.4       & 160 $\pm$ 2& 153 $\pm$ 3 & 1\\ 
  NGC 1073&1209.0&1203 $\pm$ 2&27.3&29 $\pm$ 1 &180.5$^{(2)}$& 165 $\pm$ 5& 80  $\pm$ 4 & 2\\ 
  NGC 2500&479.0 &540  $\pm$ 8&29.9&41 $\pm$ 2 & 62.7        & 85  $\pm$ 5& 111 $\pm$11 & 1\\ 
  NGC 2541&530.0 &575  $\pm$ 3&57.0&57 $\pm$ 4 & 165.0       & 172 $\pm$ 3& 96  $\pm$ 6 & 2\\ 
  NGC 2543&2469.8&2483 $\pm$ 6&59.2&61 $\pm$ 8 & 37.0        & 30  $\pm$ 3& 210 $\pm$ 3 & 1\\ 
  NGC 2712&1815.0&1858 $\pm$ 2&57.9&58 $\pm$ 5 & 3.6         & 1   $\pm$ 3& 176 $\pm$ 4 & 1\\ 
  NGC 2748&1461.3&1544 $\pm$ 6&52.9&74 $\pm$ 2 & 38.9        & 41  $\pm$ 2& 150 $\pm$ 3 & 1\\ 
  NGC 2805&1730.3&1766 $\pm$ 4&36.0&36 $\pm$ 2 & 144.2       & 123 $\pm$ 3& 106 $\pm$ 7 & 3\\ 
  NGC 3041&1400.2&1424 $\pm$ 4&50.7&50 $\pm$ 5 & 94.0        & 90  $\pm$ 5& 173 $\pm$ 7 & 1\\ 
  NGC 3403&1239.5&1282 $\pm$ 5&66.9&66 $\pm$ 4 & 74.9        & 67  $\pm$ 4& 167 $\pm$ 5 & 1\\ 
  NGC 3423&1000.8&1019 $\pm$ 4&28.4&28 $\pm$ 6 & 45.4        & 45  $\pm$ 5& 177 $\pm$ 6 & 2\\ 
  NGC 3504&1523.5&1550 $\pm$ 6&20.9&39 $\pm$ 5 & 138.3       & 165 $\pm$ 8& 151 $\pm$ 8 & 1\\ 
  NGC 4151&927.7 &1000 $\pm$ 5&48.1&21 $\pm$ 7 & 150.6       & 20  $\pm$ 3& 266 $\pm$ 4 & 4\\ 
  NGC 4324&1639.1&1689 $\pm$ 5&63.3&65 $\pm$ 2 & 55.0        & 57  $\pm$ 2& 163 $\pm$ 2 & 4\\ 
  NGC 4389&712.9 &718  $\pm$ 9&47.3&45 $\pm$ 2 & 98.9        & 100 $\pm$ 3& 149 $\pm$ 3 & 4\\ 
  NGC 4498&1656.0&1541 $\pm$ 8&57.0&58 $\pm$ 8 & 140.3       & 132 $\pm$ 7& 120 $\pm$ 8 & 2\\ 
  NGC 4639&1003.3&1025 $\pm$ 2&50.2&39 $\pm$ 4 & 128.4       & 126 $\pm$ 4& 222 $\pm$ 3 & 1\\ 
  NGC 5112&979.1 &1017 $\pm$ 5&49.2&49 $\pm$ 5 & 120.4       & 123 $\pm$ 5& 130 $\pm$ 2 & 1\\ 
  NGC 5334&1372.0&1411 $\pm$ 5&41.5&42 $\pm$ 3 & 11.9        & 10  $\pm$ 6& 166 $\pm$ 4 & 2\\ 
  NGC 5678&1898.5&1932 $\pm$ 5&56.4&60 $\pm$ 4 & 4.1         & 3   $\pm$ 3& 280 $\pm$ 4 & 1\\ 
  NGC 5740&1566.1&1600 $\pm$ 3&56.8&57 $\pm$ 5 & 162.5       & 162 $\pm$ 2& 198 $\pm$ 3 & 1\\ 
  NGC 5921&1409.4&1472 $\pm$ 3&33.2&40 $\pm$ 5 & 146.7       & 150 $\pm$ 6& 142 $\pm$ 9 & 1\\ 
  NGC 6070&2000.1&2030 $\pm$ 4&63.8&60 $\pm$ 5 & 60.0        & 57  $\pm$ 3& 231 $\pm$ 2 & 2\\ 
  NGC 6207&848.4 &869  $\pm$ 3&50.5&57 $\pm$ 7 & 19.4        & 20  $\pm$ 3& 136 $\pm$ 6 & 2\\ 
  NGC 6412&1330.4&1342 $\pm$ 2&18.4&20 $\pm$ 5 & 76.9        & 115 $\pm$ 5& 175 $\pm$ 7 & 2\\ 
  NGC 7241&1447.0&1407 $\pm$ 8&63.5&65 $\pm$ 8 & 18.5        & 23  $\pm$ 4& 142 $\pm$ 2 & 3\\ 
    \hline\end{tabular}
\end{table*}


\subsection{Non-circular motions  and residual velocity fields}
 \label{section6}

With the rotation curves we explore the circular rotation of the sample galaxies. However, there can be deviations from these circular motions provoked by dynamical features of the galaxy, such as the influence of the potential of the bar, past interactions with a companion or streaming motions across the spiral arms. To understand the influence of these features on the galaxy kinematics, we want to study these deviations from rotation, in other words, the non-circular motions.

Following the technique described in Paper I, the first step is to create a velocity model map that reflects the rotational velocities, which is done by translating the rotation curve into a 2D velocity map. The {\sc velfi} task in {\sc gipsy} was used for this, assuming kinematic symmetry and using the low-resolution rotation curve and the values of $ i $, PA, $x_{0}$, $y_{0}$ and $v_{\rm sys}$. To extract the non-circular motions we subtract the velocity model map from the observed velocity field, resulting in a residual map that is interpreted as the non-circular motions map. This residual map shows the deviations from pure rotational velocity. In panels d) and e) of Fig. \ref{ngc2748plots}, we have shown the velocity model and the non-circular motion maps respectively for NGC~2748. For the other sample galaxies, similar figures are found in the Appendix A.

Our procedure consists of studying separately those parts of galaxies most directly affected by the spiral density waves (spiral arms), those affected by the potential of the bar (bar region), { and those affected by both potentials: the start of the spiral arms close to the bar} (start-of-arms-region, SAR hereafter). We have defined these three regions (arms, bars and SAR) using images from NED (mainly SDSS false-colour images by \citealt{Baillard2011}), 3.6 $ \mu $m S$ ^{4} $G images (using the output from the 2D decompositions of the 3.6 $ \mu $m images by \citet{Salo2015}, to better constrain the bars) and our H$ \alpha $ ACAM images. Thus, we are not biased by a star formation-based classification when defining the regions, because there are barred galaxies that show the bar in the mid-IR but not in H$ \alpha $. 

{Firstly, based partly on the SDSS images but mainly on the IR images, we have defined the region where the stellar bar is, also taking into account the bar position angle and and bar length (see Sect. \ref{structural}). Secondly, we have identified by eye those regions that are between the spiral arms and the bar, and which might be influenced by both (SARs). Finally, we have identified the spiral arms by looking at the intensity contours on the SDSS and 3.6 $ \mu $m images, drawing by eye their extension on the IR images. In the more flocculent spiral arms, it is not easy to identify the extent of the spiral arms, so everything except for the bar and SAR is assumed to be spiral arms. The three regions have been defined astrometrically with irregular shapes, and can be identified at all used wavelengths. In Fig.~D1, we present the 3.6 $ \mu $m S$ ^{4} $G images of the barred galaxies with overlaid in red and yellow, a schematic representation of the bar region and the SAR for each barred galaxy. For completeness, we have added NGC~5678, a galaxy which may have a bar but which has not been classified as either SAB or SB (see Sect.~\ref{spiralncm}).}

\begin{figure}
\begin{center}
\includegraphics[width=84mm]{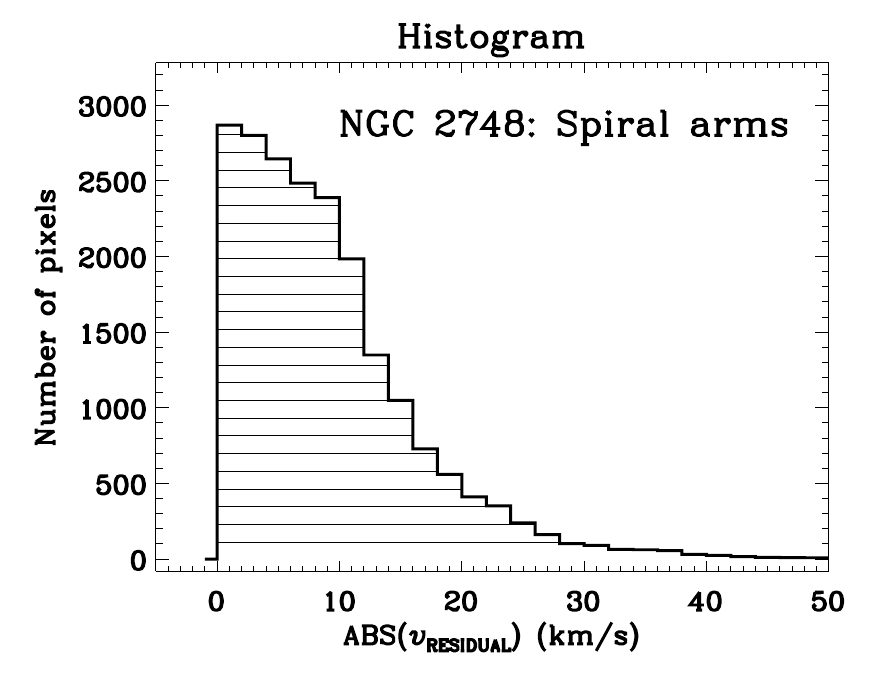}
 \includegraphics[width=84mm]{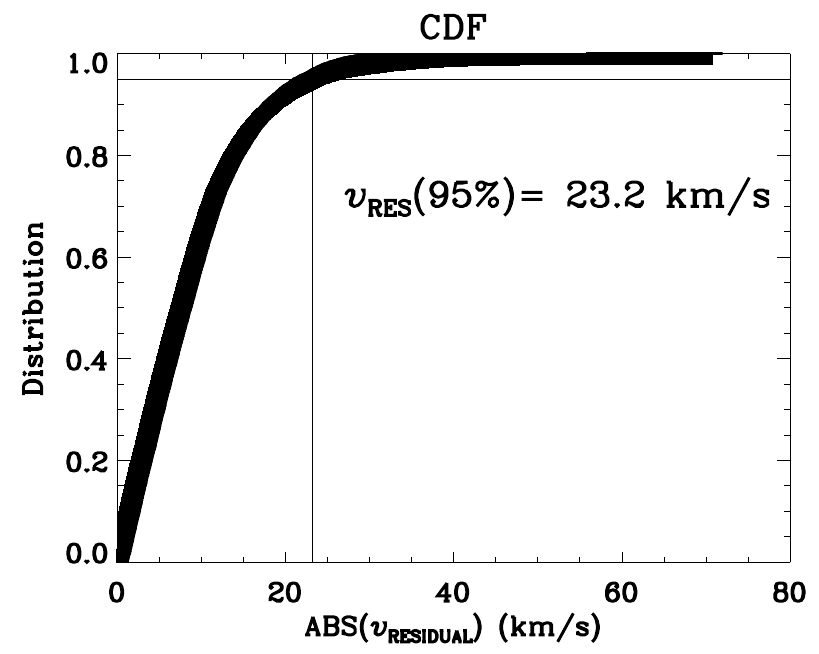}
\caption{Statistical distribution of the residual velocities of the spiral arm region of NGC~2748. \textit{Top)} Histogram distribution of the residual velocities found in the spiral arm region of the galaxy using a bin size of 1 km s$ ^{-1} $ (in absolute value). \textit{Bottom)} Cumulative distribution function of the absolute values of the residual velocities for the spiral arm region of NGC~2748. We indicate 95$\%$ value which we use as a representative overall measure of the residual velocities.}
\label{cdf}
\end{center}
\end{figure}

To quantify the non-circular motions, we have studied the cumulative distribution function (CDF) of the pixel values in the residual images. We have adopted the value of the 95$\%$ of the distribution of the residual velocities (in absolute value) as a representative value of the overall non-circular motion within that region. With this method, we avoid introducing possible caveats: if we had used a histogram, the bin size would have changed the adopted value for the residual velocity; and if we had defined regions of certain physical scale, the size of the region would matter. Another advantage of the method is that the correlations and trends would not change if we choose 50, 70 or 95$\%$, as we represent the distribution of all the values. The uncertainties were estimated via a Monte Carlo simulation of the effect of adding random noise to the velocity residuals, the former following a Gaussian distribution with sigma equal to the spectral {sampling}. In Fig. \ref{cdf} we present the statistical distribution of the residual velocities of the arm region in NGC~2748. The adopted value for the residual velocities there is 23.2 $ \pm $ 1.8 km s$ ^{-1} $; which corresponds to the representative value in the residual map of NGC~2748 (panel \textit{e} in Fig. \ref{ngc2748plots}). The computed values for the non-circular motions in the bar, SAR and spiral arms are presented in Table \ref{ncmtable}.

{For completeness, we have normalised the residual velocities. To do so, we have computed the corresponding circular velocity at a radius equal to the bar length, to the extent of SAR, and to the last measured point (assuming that the spiral arms end there), following the universal rotation curve by \citet{Persic1991}. This method includes several uncertainties, such as assuming that the circular velocity across the complete region is equal to that at the last radial point, or that the region is circular with a radius the extent of the structure (bar, SAR or arms). Note that the residual velocities have been deprojected.}

 \subsection{Structural parameters}
\label{structural}
To understand the nature of the non-circular motions within the regions of the galaxies, we study them in conjunction with some of the structural parameters that have been derived for the galaxies of our sample.  

The bar strength is quantified by the torque parameter $ Q_{\rm b} $, which describes the maximum amplitude of tangential forcing normalised by axisymmetric radial force (\citealt{Combes1981}; \citealt{Buta2001}).  $ Q_{\rm b} $ values have been measured from the torque maps derived from the 3.6 $ \mu $m S$ ^{4} $G images (for a description of the method see \citealt{Salo2010} and \citealt{Laurikainen2002}), and will be published in a compilation of bar strengths for the S$ ^{4} $G sample (S. D\'iaz-Garc\'ia et al., in prep).

The parameter $ Q_{\rm b} $ is defined so that $ Q_{\rm b} $ decreases with the axisymmetric radial force, hence the presence of the bulge is implicitly present in $ Q_{\rm b} $. Due to the bulge influence, $ Q_{\rm b} $ is not the best parameter to show the strength of the bar on the gas kinematics. However, we will use those $ Q_{\rm b} $ values as they are the only ones that measure the bar strength that have been derived from the S$ ^{4} $G images. We have also studied the bulge-to-total ratio (\textit{B/T}).  The values from \textit{B/T} have been derived from  2D decompositions of the 3.6 $ \mu $m images \citep{Salo2015}. We list the $ Q_{\rm b} $  and \textit{B/T} values for the galaxies of our sample in Table~\ref{ncmtable}. {The bar lengths have been measured from the torque maps (D\'iaz-Garc\'ia et al., in prep.). The bar lengths are deprojected units, measured using the method explained in \citet{Salo2010}, except when the fits do not have trustworthy quality, in which the bar lengths are measured visually (Herrera-Endoqui et al., in prep.).}

We have also studied the spiral arm class, as a parameter that is related to the spiral arm strength. We have obtained the arm class classifications carried out by D. Elmegreen and presented in the CVRHS, where F means \textit{flocculent}, M denotes \textit{multi-armed} and G \textit{grand design}. Note that  NGC~4324 has not been presented in the table as it does not have an arm class classification. The values for the arm classification are in Table~\ref{ncmtable}.

\begin{table*}
\caption{Non-circular motions as measured in the bar (column II), SAR (column III) and spiral arms (column IV) of the galaxies of the sample. We use the 95$\%$ value of the cumulative distribution function which we use as a representative overall measure of the residual velocities. Column V) Bar strength ($ Q_{\rm b} $) measured from the torque maps derived from the 3.6 $ \mu $m S$ ^{4} $G images (S. D\'iaz-Garc\'ia et al. in prep). Column VI) \textit{B/T} from the 2D decompositions to the 3.6 $ \mu $m images \citet{Salo2015}. Column VII) Arm class (AC) classification from {the CVRHS: F=flocculent, M=multi-armed, G=grand design}. The ``-" refers to non-barred galaxies, to those barred ones that do not have H$ \alpha $ emission within the bar or to galaxies that do not have an arm classification. {($ ^{*} $) These galaxies do not have an arm classification in the CVRHS, but as their arm classification in \citet{Elmegreen1987} was 5, we have classified them as multi-armed (M).}}
 \label{ncmtable}
\center
\begin{tabular}[angle=90]{c|clclc|clclc|}
\hline
 Galaxy name& \textit{v}$_{\rm RES,BAR}$ & \textit{v}$_{\rm RES,SAR}$ & \textit{v}$_{\rm RES,ARMS}$& $ Q_{\rm b} $ & \textit{B/T} & AC\\ 
&  (km s$ ^{-1} $) &  (km s$ ^{-1} $) &  (km s$ ^{-1} $) & & &\\
\hline
NGC 428  & 35.8 $\pm$ 1.4 &  23.5 $\pm$ 2.7 & 18.8 $\pm$ 2.7 & 0.29 $\pm$ 0.03 & 0.002 & F      \\
NGC 691  &       -        &        -        & 19.1 $\pm$ 2.5 &      -          & 0.170 & M      \\
NGC~864  & 40.1 $\pm$ 1.8 &  45.0 $\pm$ 2.4 & 17.0 $\pm$ 2.9 & 0.47 $\pm$ 0.07 & 0.027 & M      \\
NGC 918  & 24.9 $\pm$ 2.3 &  26.7 $\pm$ 0.4 & 20.2 $\pm$ 2.3 & 0.23 $\pm$ 0.02 & 0.008 & M      \\
NGC 1073 & 14.6 $\pm$ 2.4 &  13.6 $\pm$ 3.3 & 12.8 $\pm$ 3.0 & 0.63 $\pm$ 0.08 & 0.000 & M      \\
NGC 2500 & 20.6 $\pm$ 2.4 &  16.3 $\pm$ 3.5 & 14.5 $\pm$ 3.0 & 0.28 $\pm$ 0.03 & 0.002 & F      \\
NGC 2541 &       -        &        -        & 18.9 $\pm$ 2.2 &      -          & 0.000 & F      \\
NGC 2543 & 63.2 $\pm$ 0.3 &  43.2 $\pm$ 0.4 & 30.8 $\pm$ 1.7 & 0.35 $\pm$ 0.08 & 0.165 & G      \\
NGC 2712 & 48.3 $\pm$ 1.5 &  23.3 $\pm$ 1.5 & 18.5 $\pm$ 2.6 & 0.28 $\pm$ 0.05 & 0.170 & M      \\
NGC 2748 & 44.6 $\pm$ 0.4 &  21.5 $\pm$ 2.7 & 23.2 $\pm$ 1.8 & 0.45 $\pm$ 0.03 & 0.034 & M      \\
NGC 2805 & 11.1 $\pm$ 3.3 &  13.9 $\pm$ 2.6 & 15.2 $\pm$ 2.7 & 0.19 $\pm$ 0.01 & 0.002 & M      \\
NGC 3041 &       -        &        -        & 22.6 $\pm$ 2.3 &      -          & 0.043 & M      \\
NGC 3403 &       -        &        -        & 19.9 $\pm$ 1.9 &      -          & 0.000 & M      \\
NGC 3423 &       -        &        -        & 13.0 $\pm$ 3.1 &      -          & 0.055 & F      \\
NGC 3504 & 41.4 $\pm$ 2.0 &  11.4 $\pm$ 3.7 & 12.6 $\pm$ 3.4 & 0.25 $\pm$ 0.06 & 0.364 & G      \\
NGC 4151 & 11.8 $\pm$ 3.1 &  10.1 $\pm$ 3.9 &  9.6 $\pm$ 4.3 & 0.09 $\pm$ 0.02 & 0.443 & M$^{(*)}$\\
NGC 4324 &       -        &        -        & 16.2 $\pm$ 3.2 &      -          & 0.326 & -      \\
NGC 4389 & 18.6 $\pm$ 2.6 &  26.3 $\pm$ 1.5 & 21.4 $\pm$ 3.0 & 0.52 $\pm$ 0.06 & 0.000 & M$^{(*)}$\\
NGC 4498 & 19.7 $\pm$ 2.6 &  16.2 $\pm$ 2.8 & 23.9 $\pm$ 1.3 & 0.46 $\pm$ 0.07 & 0.000 & F      \\
NGC 4639 &       -        &  17.7 $\pm$ 2.5 & 15.6 $\pm$ 3.0 & 0.26 $\pm$ 0.04 & 0.112 & M      \\
NGC 5112 & 14.0 $\pm$ 2.9 &  23.4 $\pm$ 2.7 & 19.9 $\pm$ 2.1 & 0.62 $\pm$ 0.06 & 0.000 & M      \\
NGC 5334 &       -        &  18.1 $\pm$ 2.9 & 15.3 $\pm$ 2.8 & 0.49 $\pm$ 0.08 & 0.001 & F      \\
NGC 5678 &       -        &        -        & 72.6 $\pm$ 0.1 &      -          & 0.037 & F      \\
NGC 5740 &       -        &  21.0 $\pm$ 2.4 & 21.1 $\pm$ 1.7 & 0.16 $\pm$ 0.03 & 0.125 & M      \\
NGC 5921 &       -        &  21.9 $\pm$ 1.7 & 16.5 $\pm$ 2.6 & 0.34 $\pm$ 0.06 & 0.111 & M      \\
NGC 6070 &       -        &        -        & 20.7 $\pm$ 1.7 &      -          & 0.045 & M      \\
NGC 6207 & 18.6 $\pm$ 2.6 &  15.5 $\pm$ 3.0 & 17.9 $\pm$ 2.3 & 0.21 $\pm$ 0.02 & 0.000 & F      \\
NGC 6412 & 12.7 $\pm$ 2.9 &   5.2 $\pm$ 6.7 & 15.0 $\pm$ 2.6 & 0.24 $\pm$ 0.02 & 0.005 & M      \\
NGC 7241 &       -        &        -        & 18.6 $\pm$ 2.3 &      -          & 0.000 & M      \\
 \hline\end{tabular}
\end{table*}


\subsection{Star formation rates}
 \label{section7}
 
The H$ \alpha $ line traces the emission from massive young stars. {The advantages of this recombination line are that it traces very recent SF (timescales $ \sim $6-8 Myr), and that it has lower sensitivity to dust attenuation than the UV (although not negligible, \citealt{Cardelli1989}).} Thus, we can exploit the H$ \alpha $ imaging and relate the star formation to the galaxy's kinematics. We derive the star formation rates (SFRs) derived from the H$ \alpha $ luminosity (L$ _{\rm H\alpha} $) following \citet{Kennicutt2009}:
 
 \begin{equation}
 \label{sfrha}
 {\rm SFR} (M_{\odot}  {\rm yr^{-1}}) = 5.5 \times 10^{-42} L({\rm H\alpha}),
 \end{equation}
 
 where $L({\rm H\alpha})$ is the luminosity, calculated as
 
 \begin{equation}
 L({\rm H\alpha}) [{\rm erg/s}]=4\pi D^{2} (3.086 \times 10^{24})^{2}F_{{\rm H\alpha}}^{*},
\end{equation}  

with \textit{D} the distance to the galaxy in Mpc (Table 1) and $F_{{\rm H\alpha}}^{*}$ the flux corrected for Galactic absorption, as taken from NED and obtained from the \citet{Schlafly2011} recalibration of the \citet{Schlegel1998} dust map. {Equation \ref{sfrha} assumes a \citet{Kroupa2001} stellar IMF with a mass range of 0.1-100 $M_{\odot} $,  an electron temperature of $ T_{\rm e} =10^{4} $ K and electron density $ n_{\rm e} =100$ cm$ ^{-3} $, whereas variations in $ T_{\rm e} $ from 5000 to 20000 K would result in a variation of the calibration constant ($5.5 \times 10^{-42}$) of $\sim$15\%. Variations of $ n_{\rm e}  = 100-10^{6} $ cm$ ^{-3} $ would result in variations in the calibration constant below 1\% \citep{Osterbrock2006}. This calibration also assumes that over timescales  $>$6 Myr, star formation needs to remain constant, and no information about the previous star formation history is given (see \citealt{Kennicutt2012}).}

The uncertainties in the measurements of the SFR come from: mainly from the distance, with a typical 20\% uncertainty estimated for the distance value, basic reduction processes (i.e. flat-fielding correction, around 2\%), the zero-point calibration (3\%, in agreement with the value of 2\% typical for photometric nights), uncertainty in the flux measurement (1\%) and continuum-subtraction process (around 11\%). Altogether, taking into account the uncertainties related to the quality of the image (without considering the distance) we estimate an uncertainty of 17\%.

To consider only star formation, we have to take into account that some of the galaxies of our sample have nuclear activity: NGC~918 (no specific classification; \citealt{Palumbo1983}); NGC~4151 (Seyfert type 1.5; \citealt{Khachikian1974}) and NGC~4639 (Seyfert 1.0; \citealt{Ho1997}), where all nuclear activity classifications have been collected in \citet{VeronCetty2006}. The H$ \alpha $ emission in these cases is not all coming from star formation, so the derived SFRs in these galaxies should be understood as upper limits. We have masked the nuclear emissions in these three cases and excluded them from the integration (very small in the cases of NGC~918 and the low-luminosity AGN of NGC~4639), so that the global H$ \alpha $ luminosities are not biased due to AGNs. NGC~3504 on the other hand is a starburst galaxy \citep{Balzano1983} with a large amount of star formation in its centre. 

\begin{figure}
\begin{center}
 \includegraphics[width=84mm]{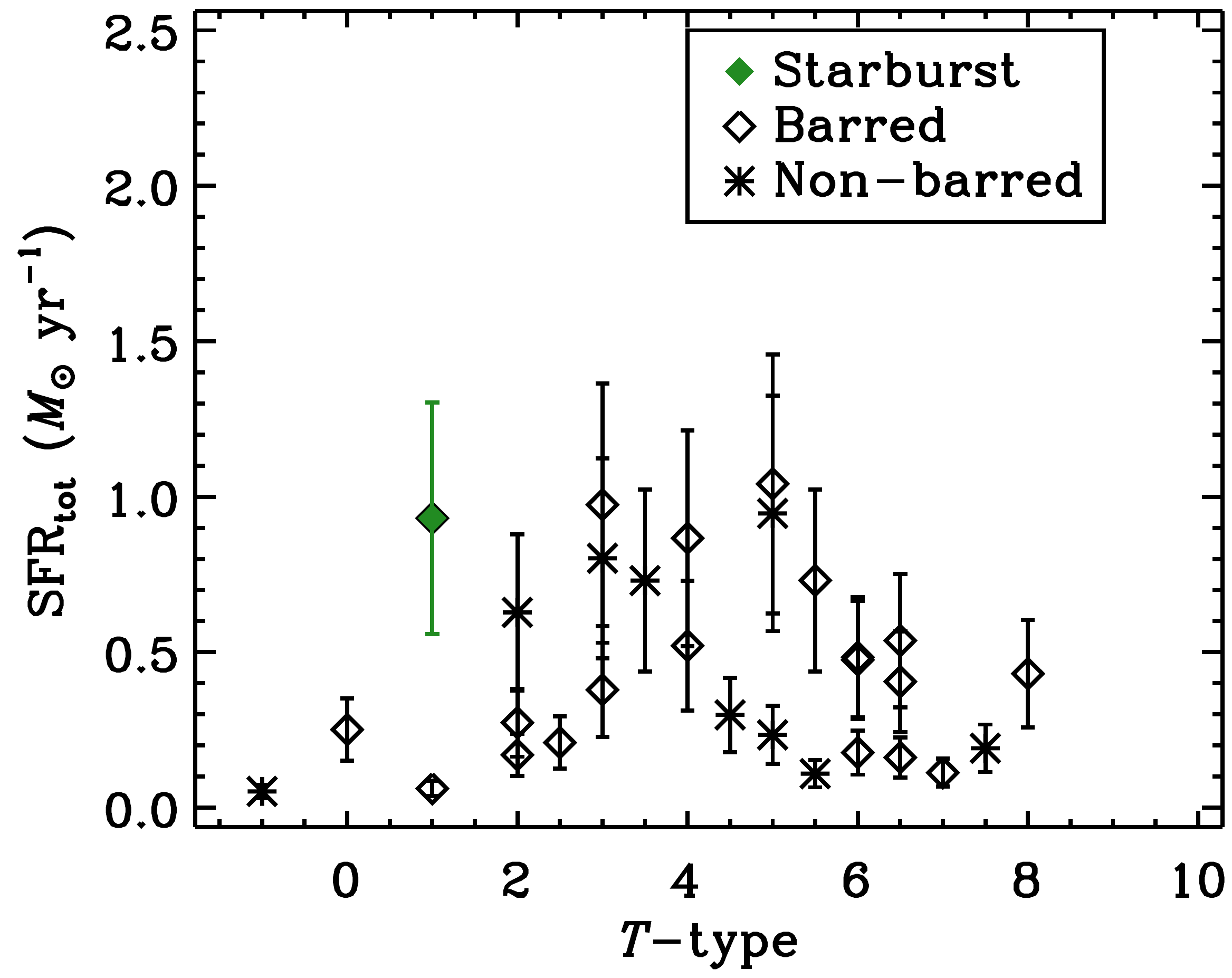}
\caption{SFR as a function of morphological \textit{T}-type. Barred galaxies are presented with diamonds, whereas non-barred galaxies are presented with asterisks. The starburst galaxy NGC~3504 is identified with a filled green diamond.}
\label{sfr}
\end{center}
\end{figure}

In Fig. \ref{sfr}, we present the SFR as a function of the morphological \textit{T}-type (from {the CVRHS}). Excluding the starburst galaxy (which stands out in the plot as a filled green diamond), we find the same trend as widely found in the literature  (e.g., \citealt{Kennicutt1998}; \citealt{James2004}; and references therein): early-type galaxies tend to have low SFRs, intermediate-type spirals have the highest SFRs and very late-type spirals and irregulars have lower SFRs than the intermediate-type ones, although higher than those of galaxies of the earliest types. Although our sample is small, we confirm that the presence of a bar does not affect the global SFR (e.g., \citealt{Phillips1993}; \citealt{Kennicutt1994}; \citealt{Kennicutt1998} and references therein; \citealt{James2004}; \citealt{Fisher2006}). {A Kolmogorov-Smirnov (K-S) test on both the barred and the non-barred distributions with respect to the SFR results in a \textit{P}-value of 0.094. This \textit{P}-value indicates the relationship between both distributions, and in this case, it means that both the barred and the non-barred distributions with respect to the SFR are very similar.

From our ACAM images, we have measured the total SFRs for the galaxies of our sample, and also the SFRs for each of the defined regions. In Table~\ref{sfrtable} we present the resulting SFRs (total, bar, SAR and arms).

}

\begin{table*}
\caption{SFRs measured from the ACAM H$ \alpha $ images, corrected for Galactic absorption (see text). \textit{Notes}:  The ``-" refers to non-barred galaxies or to those barred ones that do not have H$ \alpha $ emission within the bar. (*) The AGN contribution has been excluded from the measurements.}
 \label{sfrtable}
\center
\begin{tabular}[angle=90]{|c|c|c|c|c|}
\hline
 Galaxy name& SFR$ _{\rm tot} $ &SFR$ _{\rm bar} $ &SFR$ _{\rm SAR} $ &SFR$ _{\rm arms} $  \\ 
  &  ($M_{\odot} $ yr$ ^{-1} $) & ($M_{\odot} $ yr$ ^{-1} $)& ($M_{\odot} $ yr$ ^{-1} $)& ($M_{\odot} $ yr$ ^{-1} $)\\
\hline
 NGC  428 &0.431        $\pm$ 0.172 & 0.019 $\pm$  0.008 & 0.078 $\pm$  0.031 & 0.334 $\pm$  0.134 \\
 NGC  691 &0.628        $\pm$ 0.251 &         -          &         -          & 0.628 $\pm$  0.251 \\
 NGC  864 &0.867        $\pm$ 0.347 & 0.082 $\pm$  0.033 & 0.035 $\pm$  0.014 & 0.750 $\pm$  0.300 \\
 NGC  918 &0.475$^{(*)}$ $\pm$0.190 & 0.006  $\pm$ 0.002 & 0.004  $\pm$ 0.002 & 0.465  $\pm$ 0.186 \\
 NGC 1073 &0.731        $\pm$ 0.293 & 0.029 $\pm$  0.011 & 0.018 $\pm$  0.007 & 0.685 $\pm$  0.274 \\
 NGC 2500 &0.161       $\pm$ 0.065 & 0.002 $\pm$  0.001 & 0.006 $\pm$  0.002 & 0.154 $\pm$  0.062 \\
 NGC 2541 &0.191        $\pm$ 0.076 &         -          &         -          & 0.191 $\pm$  0.076 \\
 NGC 2543 &0.379        $\pm$ 0.152 & 0.060 $\pm$  0.024 & 0.022 $\pm$  0.009 & 0.297 $\pm$  0.119 \\
 NGC 2712 &0.209        $\pm$ 0.084 & 0.063 $\pm$  0.025 & 0.021 $\pm$  0.008 & 0.125 $\pm$  0.050 \\
 NGC 2748 &0.521        $\pm$ 0.208 & 0.018 $\pm$  0.007 & 0.158 $\pm$  0.063 & 0.345 $\pm$  0.138 \\
 NGC 2805 &1.041        $\pm$ 0.417 & 0.003 $\pm$  0.001 & 0.017 $\pm$  0.007 & 1.022 $\pm$  0.409 \\
 NGC 3041 &0.730        $\pm$ 0.292 &         -          &         -          & 0.730 $\pm$  0.292 \\
 NGC 3403 &0.234        $\pm$ 0.094 &         -          &         -          & 0.234 $\pm$  0.094 \\
 NGC 3423 &0.298        $\pm$ 0.119 &         -          &         -          & 0.298 $\pm$  0.119 \\
 NGC 3504 &0.931        $\pm$ 0.373 & 0.687 $\pm$  0.275 & 0.148 $\pm$  0.059 & 0.097 $\pm$  0.039 \\
 NGC 4151&0.251$^{(*)}$ $\pm$ 0.100 & 0.025 $\pm$  0.010 & 0.009 $\pm$  0.003 & 0.217 $\pm$  0.087 \\
 NGC 4324 &0.052        $\pm$ 0.021 &         -          &         -          & 0.052 $\pm$  0.021 \\
 NGC 4389 &0.061        $\pm$ 0.025 & 0.035 $\pm$  0.014 & 0.013 $\pm$  0.005 & 0.013 $\pm$  0.005 \\
 NGC 4498 &0.112        $\pm$ 0.045 & 0.011 $\pm$  0.004 & 0.021 $\pm$  0.008 & 0.081 $\pm$  0.032 \\
 NGC 4639 &0.170$^{(*)}$ $\pm$0.068 & 0.003  $\pm$ 0.001 & 0.045  $\pm$ 0.018 & 0.122  $\pm$ 0.049 \\
 NGC 5112 &0.537        $\pm$ 0.215 & 0.020 $\pm$  0.008 & 0.099 $\pm$  0.040 & 0.418 $\pm$  0.167 \\
 NGC 5334 &0.177        $\pm$ 0.071 & 0.002 $\pm$  0.001 & 0.018 $\pm$  0.007 & 0.157 $\pm$  0.063 \\
 NGC 5678 &0.802        $\pm$ 0.321 &         -          &         -          & 0.802 $\pm$  0.321 \\
 NGC 5740 &0.273        $\pm$ 0.109 & 0.005 $\pm$  0.002 & 0.062 $\pm$  0.025 & 0.206 $\pm$  0.083 \\
 NGC 5921 &0.974        $\pm$ 0.390 & 0.014 $\pm$  0.006 & 0.107 $\pm$  0.043 & 0.853 $\pm$  0.341 \\
 NGC 6070 &0.946        $\pm$ 0.379 &         -          &         -          & 0.946 $\pm$  0.379 \\
 NGC 6207 &0.405        $\pm$ 0.162 & 0.007 $\pm$  0.003 & 0.113 $\pm$  0.045 & 0.285 $\pm$  0.114 \\
 NGC 6412 &0.484        $\pm$ 0.194 & 0.030 $\pm$  0.012 & 0.002 $\pm$  0.001 & 0.452 $\pm$  0.181 \\
 NGC 7241 &0.109        $\pm$ 0.044 &         -          &         -          & 0.109 $\pm$  0.044 \\
       \hline\end{tabular}
\end{table*} 


\section{Data release}
 \label{section8}
 
All the data discussed here are released publicly with the publication of this paper: raw but reduced cubes {(de-rotated, phase corrected and wavelength calibrated cubes)}, continuum-subtracted cubes, 0th, 1st and 2nd order moment maps and continuum-subtracted H$ \alpha $ images. The data are available  in FITS format through the NED and the Centre de Donn\'ees Stellaires (CDS).

\section{Discussion}
 \label{section9}

In this Section we will discuss critically the data  and data quality, and the results from our analysis in terms of kinematical parameters and non-circular motions. We will not analyse here the rotation curves as they will be studied in depth in the forthcoming Paper III (Erroz-Ferrer et al., in prep.).

\subsection{Data quality}

We first discuss how our observing, reduction and analysis procedures have led to caveats and constraints on the data quality. 

\subsubsection{Effects from the nature of the H$ \alpha $ data}

One of the scientific goals of this kinematical survey is the study of the inner parts of the rotation curves, exploiting the high angular resolution provided by GH$ \alpha $FaS data. To do so, we have tried not to change the angular resolution when reducing the data, as would have been the case when using large spatial smoothing kernels or adaptive binning methods. As the data are made public, the user can always choose to adopt another smoothing kernel depending on scientific interest.

Normally, H$ \alpha $ data have a typical patchy appearance with many blank spaces in between the H{\sc ii} regions. As we have hardly smoothed the data spatially, this effect is rather obvious in most of our data. The data set contains  galaxies with high signal and less patchy appearance (such as NGC~2748 or NGC~5678), and also galaxies with low signal and patchier appearance (e.g., NGC~691 or NGC~918). Consequently, the process of deriving rotation curves is difficult and leads to unsampled radial regions or bad radial sampling. For example, the central regions of the galaxies NGC~428, NGC~4151, NGC~4324, NGC~4639, NGC~5334, NGC~5740 and NGC~5921 are not sampled and can not be studied.

\subsubsection{Kinematical parameters versus photometric parameters}

\begin{figure}
\begin{center}
 \includegraphics[width=84mm]{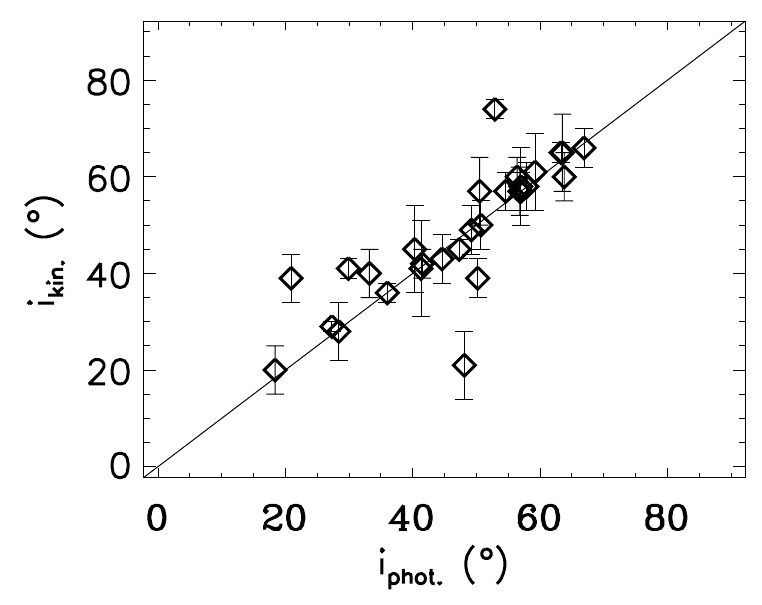}
 \includegraphics[width=84mm]{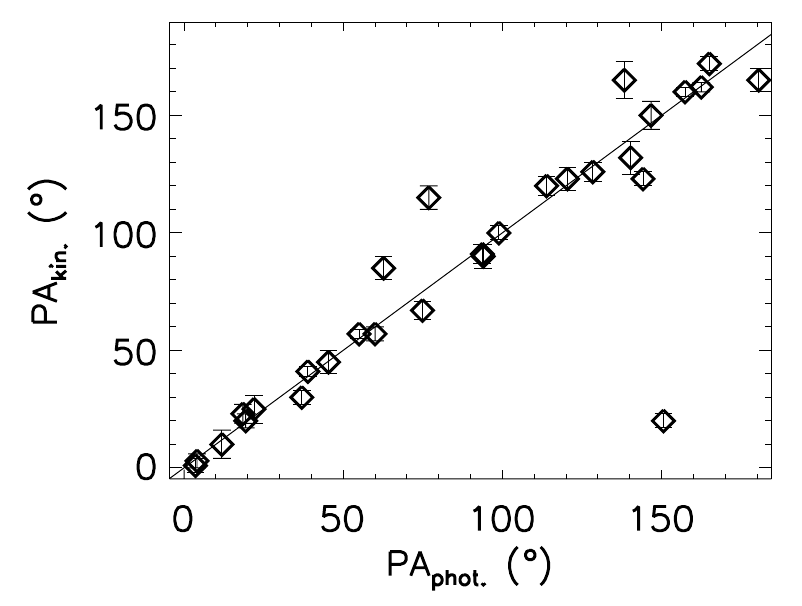}
\caption{Comparison between the parameters inclination and position angle derived from the S$ ^{4} $G 3.6 $ \mu $m images (i.e. photometric) and derived with the tilted-ring method to our H$ \alpha $ FP data (i.e. kinematical). Left) Comparison of the inclinations derived photometrically and kinematically. Right) Comparison between the position angles derived photometrically and kinematically. \textit{Note:} Due to our definition of the position angles ($0\degr \leq \rm PA < 180\degr$), the difference between PA=0$\degr$ and PA=179$\degr$ is not 179$\degr$, but 1$\degr$.}
\label{comparisons}
\end{center}
\end{figure}

Another caveat (though related to the first point) is the determination of the kinematical parameters $ i $, PA or $v_{\rm sys} $. We explained in Sect. \ref{section5} how we have derived our rotation curves using  {\sc rotcur} in GIPSY. Due to the nature of our data, it is not easy to determine the kinematical parameters, and it is more often reliable to use other types of data, with full spatial coverage and full spatial sampling albeit at lower angular resolution. Thus, in our fits we have used H{\sc i}-derived parameters, whenever possible, as initial conditions. We also use the photometric information provided by the ellipse fitting analysis on the S$ ^{4} $G 3.6 $ \mu $m images by Mu\~noz-Mateos et al. (in prep), although our results do not always match the photometric results (see Table \ref{rotcuroutput} for a comparison).

Regarding the PA, the PV diagrams and the position of the minor axis in our velocity maps are good tracers of the kinematical PA, so the results from {\sc rotcur} describe the gas motion better than the photometric results. There is one case with a high discrepancy: NGC~4151 with PA$_{3.6} $=150.6$\degr$ and PA$_{\rm H\alpha} $=20$\degr \pm$ 3$\degr$ (the real difference is the same as  with PA$_{\rm H\alpha} $=180$\degr$+20$\degr$=200$\degr$, that is 49.4$\degr$). After an inspection of the continuum-subtracted cube, and also by looking at the velocity map (panel b of Fig. A8 bottom) we have found that the PA is closer to 0$\degr$, very different from the 150$\degr$ derived from the S$ ^{4} $G image. \citet{Bosma1977} discussed the first H{\sc i} synthesis maps of this galaxy, along with a deep photograph by Arp showing the outer spiral. They found that the PA$_{\rm kin} $=19$\degr$ $\pm$ 4$\degr$, much closer to our PA$_{\rm H\alpha} $ than to PA$_{\rm phot} $. We confirm Bosma's conclusion: the photometric image detects only the oval distribution of the galaxy.

In the same way, there are other galaxies in the sample where the mid-IR morphology results in PA$_{\rm phot}$ different from PA$_{\rm kin} $, but our PA$_{\rm kin} $ agree with previous values of PA$_{\rm kin} $ found in the literature. {These are presented in Table~\ref{pacomparison}. All the galaxies in Table~\ref{pacomparison} are barred (NGC~2805 is classified as SA in the CVRHS, but is claimed to host a bar in the 2D structural decomposition of the S$ ^{4} $G images, \citealt{Salo2015}). Therefore, the differences between the photometric and kinematic PA may well be explained with the presence of the bar which, depending on its orientation, influences the kinematics in the bar region and causes deviations from pure circular motions (see Sect.~\ref{ncm}). Also, irregular outer spiral arms (such in NGC~6412) can influence on the measurements of PA$_{\rm phot}$, and we conclude that PA$_{\rm kin}$ trace the real PA better than PA$_{\rm phot}$.}

\begin{table*}
\caption{PA from different sources, measured in degrees and defined from North to East. \textit{Column I)} PA$_{\rm phot}$ from the ellipse fitting to the 3.6 $ \mu $m S$ ^{4} $G images. \textit{Column~II)} PA$_{\rm kin}$ from our H$ \alpha $ FP data. \textit{Column~III)} PA$_{\rm kin}$ values found in the literature. \textit{Column~IV)} References.}
 \label{pacomparison}
\center
\begin{tabular}{|c|c|c|c|c|}
\hline
Galaxy & PA$_{\rm phot} $& PA$_{\rm kin,THIS~STUDY} $ &  PA$_{\rm kin,lit} $ & Reference\\    
\hline
NGC~1073 & 180.5 & 165 $\pm$ 4 & 164.6 & \citet{England1990}\\ 
NGC~2500 & 62.7 & 85 $\pm$ 5 & 85 &  GHASP~VII \\ 
NGC~2805 &144.2 & 123 $\pm$ 3 & 114 &  GHASP~VII\\ 
NGC~3504 &138.3 & 165 $\pm$ 8 & 163 &  GHASP~VII \\ 
NGC~6412 & 76.9 & 115 $\pm$ 5 & 115 &  GHASP~VII \\ 
\hline   
  \end{tabular}
\end{table*}

As for the inclination, the largest discrepancies occur in NGC~4151 (already discussed), and in NGC~2748 and NGC~3504, where the inclinations as derived from the kinematic data are substantially higher than those derived from the surface photometry. This can be due to the non-uniform distribution of the H{\sc ii} regions as compared to the distribution of the stars and dust as seen in the S$ ^{4} $G 3.6 $ \mu $m images. {NGC~2748 shows a bulge at faint light levels, hence the discrepancy in inclination. NGC~3504 has a low inclination, seen in the S$ 4 $G image}. Also, NGC~2748 and NGC~3504 host an outer ring (R' and R'$ _{1} $, respectively), not seen in H$ \alpha $.

\begin{figure*}
\begin{center}
 \includegraphics[width=175mm]{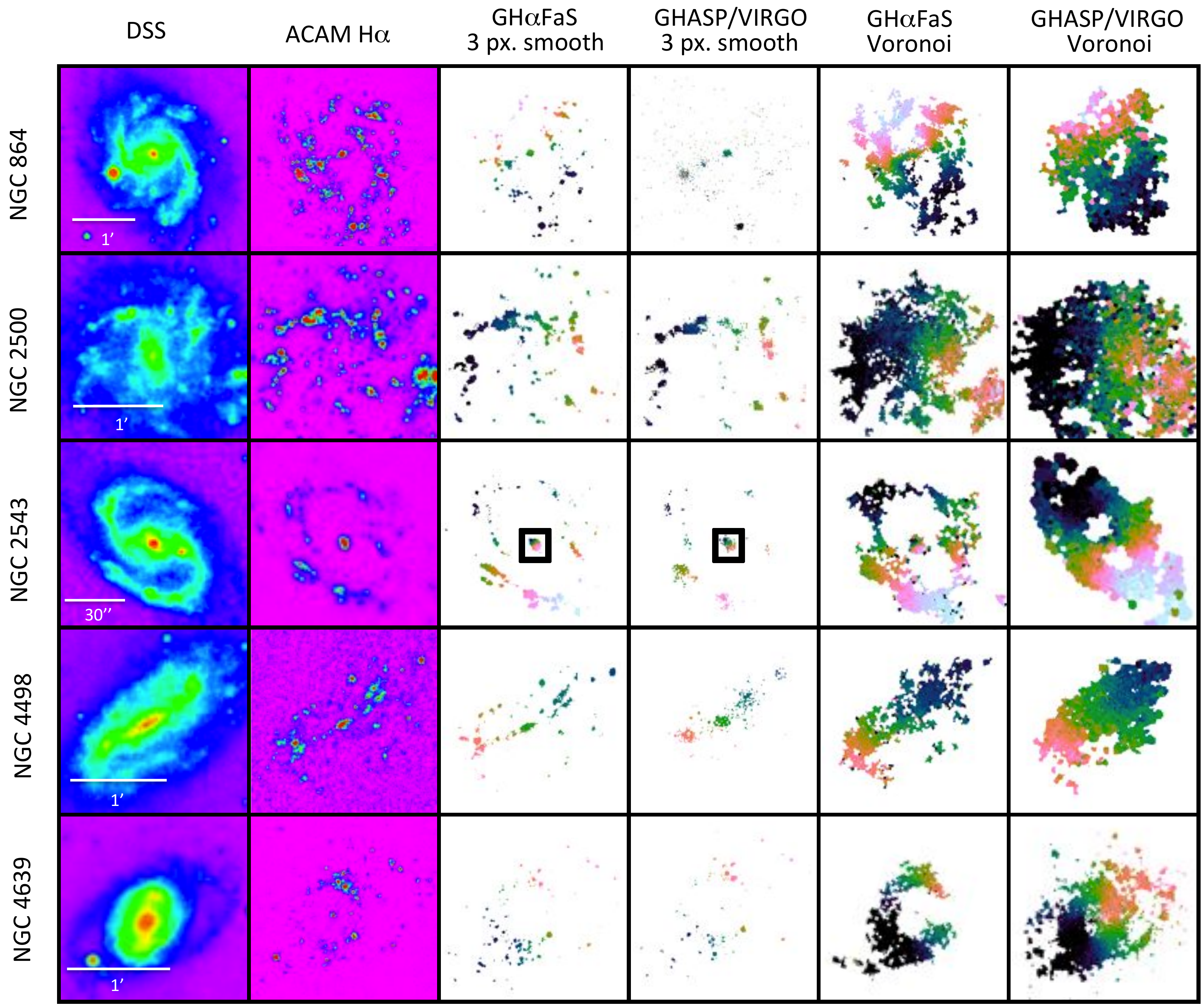}
\caption{Comparison between the data obtained with the GH$ \alpha $FaS instrument and data from GHASP (NGC~864, NGC~2500 and NGC~2543) {and VIRGO (NGC~4498 and NGC~4639)}. Column I) DSS image. Column~II) ACAM H$ \alpha $ image. Column~III) Velocity field from GH$ \alpha $FaS observations, obtained following the reduction procedures described in Sect. \ref{section4}. {Column~IV) Velocity field from GHASP/VIRGO observations, obtained following the reduction procedures described in Sect. \ref{section4}.} Column~V) Velocity field from GH$ \alpha $FaS observations following the reduction processes of GHASP{/VIRGO} data (three-channel spectral smoothing and Voronoi adaptive binning). Column~VI) Data from GHASP {\citep{Epinat2008} and VIRGO \citep{Chemin2006} surveys}. {\textit{Note)} For NGC~2543, the central region has been marked with a black square and explicitly shown in Fig~\ref{comparisoghasp2}.}}
\label{comparisoghasp}
\end{center}
\end{figure*}

{
\subsubsection{Comparison with literature data}

In the Introduction, we mentioned the H$ \alpha $ FP kinematic surveys of \citealt{Garrido2002} (GHASP),  \citealt{Daigle2006a} (SINGS sample) and \citealt{Chemin2006} (VIRGO sample). These surveys have used instrumentation similar to GH$ \alpha $FaS: Cigale and  FaNTOmM (Fabry–Perot de Nouvelle Technologie de l'Observatoire du Mont M\'egantic\footnote{http://www.astro.umontreal.ca/fantomm}). These were mounted at {different} telescopes: the 3.6 m Canada-France-Hawaii Telescope (VIRGO and SINGS), {the 3.6 m European Southern Observatory telescope (VIRGO)}, the 1.93 m Observatoire de Haute-Provence telescope (VIRGO, SINGS and GHASP), and the 1.60 m Observatoire du Mont M\'egantic telescope (VIRGO and SINGS). All these data are seeing-limited, but taken under observing conditions worse than ours (less than 2" seeing in the very best cases, but with seeing values of 8"-12" in the worst cases). In Table~\ref{tablecomparison}, we show the different quality characteristics of these surveys as compared to our GH$ \alpha $FaS data. 

\begin{table*}
\caption{Characteristics of the H$ \alpha $ FP kinematical surveys of \citealt{Garrido2002} (GHASP),  \citealt{Daigle2006a} (SINGS sample) and \citealt{Chemin2006} (VIRGO sample), shown along with the characteristics of the data used in this paper. \textit{Note)} The `-' denotes that there are no galaxies with that criteria (VIRGO, GHASP and this paper), or that no information about the seeing was presented (SINGS). $ ^{(*)} $ Information before the reduction processes (i.e., before smoothing).}
 \label{tablecomparison}
\center
\begin{tabular}{|c|c|c|c|c|}
\hline
Parent survey & VIRGO & SINGS & GHASP &  S$ ^{4} $G (this paper)\\
\hline
 Sample size & 30 & 28 & 203 & 29\\    
\hline
 Seeing $ < 1\farcs5 $ &  11 &-  & -  &29\\
 2" $ < $ Seeing $ \lesssim $ 4" & 19 &- & 143  & - \\
 4" $ < $ Seeing $ \lesssim $ 6" &-  & -& 45 & - \\
 Seeing $ > 6\farcs0 $  &- &- & 15& - \\
\hline
 Angular sampling & 0\farcs42 - 1\farcs61 &0\farcs42 - 1\farcs61 & 0\farcs68 - 0\farcs96 & 0\farcs2 \\
\hline
 Spatial smoothing & Voronoi S/N=5 &Voronoi S/N=5 &  Voronoi S/N=5  &Median 3$ \times $3 pix\\
 & Gaussian 3"$ \times $3" & & & \\
\hline
 Spectral sampling$ ^{(*)} $& 7 - 14  km s$ ^{-1} $ &7 - 14  km s$ ^{-1} $ & $\sim$10km s$ ^{-1} $ & $\sim$8 km s$ ^{-1} $\\ 
\hline     \end{tabular}
\end{table*}

The reduction techniques used in these surveys are based on the University of Montr\'eal Improved 3D Fabry-Perot Data Reduction Techniques\footnote{http://www.astro.umontreal.ca/~odaigle/reduction/} and are summarised in \citet{Daigle2006}. The most important reduction steps that differ from ours are spectral and spatial smoothing. The three surveys smooth the data spectrally (that is, in the wavelength direction), with three-channel or Hanning-filtering smoothing. Also, they use the Voronoi adaptive binning method \citep{Cappellari2003}, which consists of gathering as many pixels as necessary in order to achieve a threshold value of the signal-to-noise ratio for the considered bin. Hence, the Voronoi method allows maximum spatial resolution for regions with strong emission and good signal-to-noise values for regions with low emission.

On the contrary and with the goal of keeping the highest spectral and angular resolution possible throughout, we have not smoothed the data spectrally and applied a median spatial smoothing of 3$\times$3 pixels. These reduction procedures have been chosen following our scientific goals of studying the inner parts of the rotation curves and studying streaming motions within the different galaxy substructures. 

In Fig. \ref{comparisoghasp},  {we compare the velocity fields of three galaxies in the overlap of our sample with that of GHASP (NGC~864, NGC~2500 and NGC~2543) and two galaxies in the overlap with the VIRGO sample (NGC~4498 and NGC~4639). We compare the velocity fields as obtained from the GH$ \alpha $FaS, GHASP and VIRGO data by different reduction methods: ours, as explained in Sect. \ref{section4}, and following the reduction techniques used for the GHASP and VIRGO surveys (three-channel spectral smoothing and Voronoi adaptive binning, see \citealt{Chemin2006} and \citealt{Epinat2008})}. In these figures, it is possible to compare directly the advantages and disadvantages of each reduction procedure. On the one hand, we see that one drawback of following the reduction steps of Sect. \ref{section4} is that the spatial coverage is limited to the location of the H{\sc ii} regions, resulting in patchy maps. But on the other hand, Voronoi smoothing reduces the angular resolution often to that typical of H{\sc i} data (6"-10"). With the results from the adaptive binning, it is very difficult to perform studies inside the structures of the galaxies (such as bars or spiral arms), as they cannot be resolved any more. This includes the study of the streaming motions (this paper), and of the inner parts of the rotation curves (Paper III). As explained in Sect.~\ref{section8}, we publish our raw and basic reduced data, and any worker interested in smoothed versions of our data sets, either spectrally or Voronoi, can simply download the data and perform their own favourite recipes.}

\begin{figure}
\begin{center}
 \includegraphics[width=84mm]{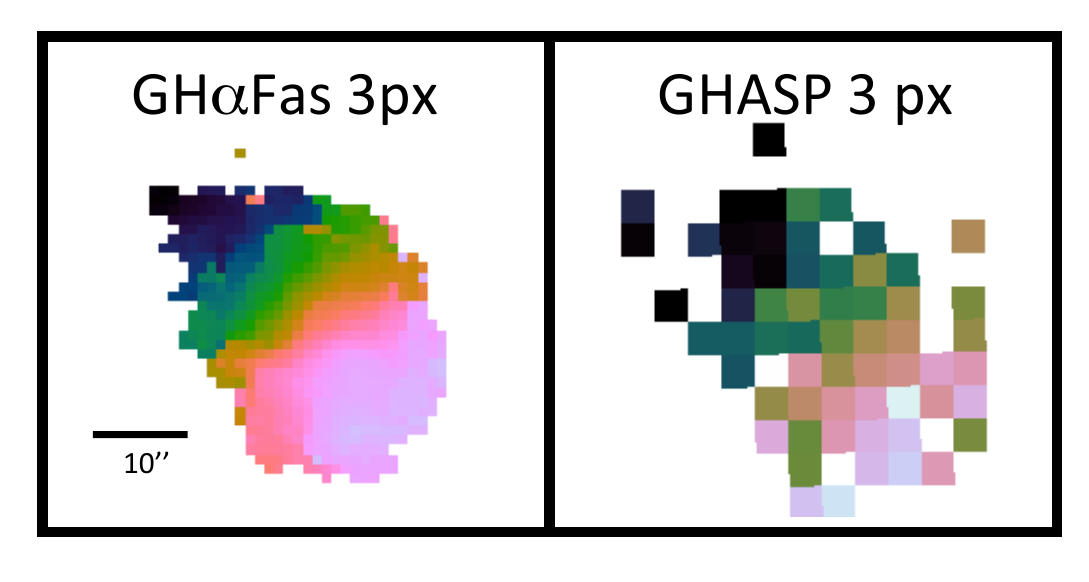}
\caption{{Velocity fields of the central region of NGC~2543 reduced following the procedures of this paper from GH$ \alpha $FaS (left) and GHASP (right) observations. Note the differences due to the pixel scale (0$\farcs$2 of GH$ \alpha $FaS against 0$\farcs$68 of GHASP) and seeing conditions ($1\farcs3$ for our observations against $3\farcs7$ for GHASP).}}
\label{comparisoghasp2}
\end{center}
\end{figure}

{In Fig.~\ref{comparisoghasp2} we present the velocity field of the central part of NGC~2543 to show the differences between the GH$ \alpha $FaS and GHASP data, both reduced following the techniques explained in the current paper. The differences arising from the different pixel scale (0$\farcs$2 of GH$ \alpha $FaS against 0$\farcs$68 of GHASP) and different seeing conditions ($1\farcs3$ for our observations against $3\farcs7$ for GHASP) can clearly be seen.}

\begin{figure*}
\begin{center}
 \includegraphics[width=150mm]{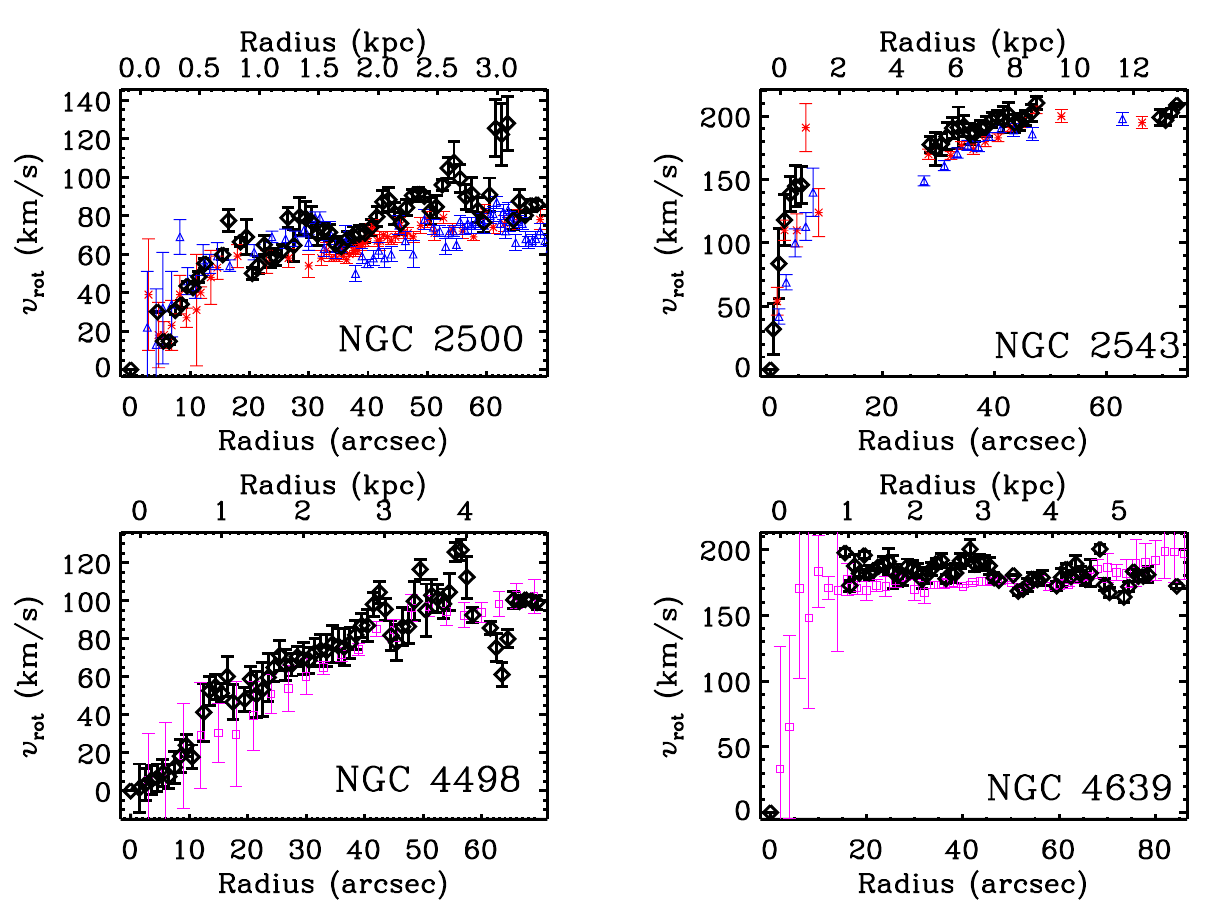}
\caption{{H$ \alpha $ rotation curves for NGC~2500, NGC~2543, NGC~4498 and NGC~4639. Black diamonds correspond to the high-resolution rotation curves derived from GH$ \alpha $FaS data (this paper). Red asterisks and blue triangles correspond to approaching and receding rotation curves from GHASP data. Pink squares correspond to VIRGO data.}}
\label{rotcurcomparison}
\end{center}
\end{figure*}

{To see the possible impact of the different reduction processes, we also compare the rotation curves of the overlapping galaxies. In Fig.~\ref{rotcurcomparison} we superimpose the rotation curves from GHASP and VIRGO papers onto our rotation curves for the galaxies in Fig.~\ref{comparisoghasp}. \citet{Erroz-Ferrer2012} give a direct comparison between the rotation curve of NGC~864 from our data and that of GHASP, so this galaxy has not been included in Fig.~\ref{rotcurcomparison}. For completeness, we compare the rotation curves of the remaining overlapping galaxies in Figs.~B1-B4. There are many similarities and the rotation curves agree reasonably.  The possible discrepancies mainly come from the different parameters ($v_{\rm sys}$, $i$ and PA) used for the fits to the velocity fields. Also, the spatial coverage resulting from the different reduction techniques is reflected in the rotation curves: smoothing kernels such as Voronoi help to highlight the emission in the low-signal regions, but compromise the spatial resolution.}

\subsection{Non-circular motions}
\label{ncm}

One of the aims of this kinematical study is to analyse the influence of the galaxies' main structural features on their kinematics. In other words, we want to study the kinematical footprints of the components of the galaxies, traced for instance by the non-circular motions. Deviations from pure circular motion have been widely studied previously in the literature. The first studies of streaming motions induced by the spiral density waves were presented by \citet{Bosma1978}, \citet{Visser1980} and \citet{Rots1990} using H{\sc i} data; by \citet{Ichikawa1985}, \citet{Clemens1985} and \citet{Cepa1992} using CO data, and by  \citet{vanderKruit1976} and \citet{Marcelin1985} using H$ \alpha $ data. {They confirm what \citet{Roberts1969b} had predicted theoretically: variations of $10-30$ km s$^{-1} $ in the velocity are found when the gas crosses the density wave.} Bar-induced non-circular motions have also been analysed in the literature, and the first studies were performed by \citet{Peterson1978}, \citet{Duval1983} and \citet{Pence1988} using optical data; and by \citet{Sancisi1979} and \citet{Gottesman1984} using H{\sc i} data. {They found that the isovelocity contours tend to go parallel to the bar rather than parallel to the minor axis. Also, they found some cases where these deviations from the circular motion are more prominent when the PA$ _{\rm BAR} $ is at $\sim45\degr$ from the kinematic major axis (in agreement with the models by, e.g., \citealt{vanAlbada1981}). The nature and characterisation of these non-circular motions is a topic still under debate, and many studies of the streaming motions have been performed (e.g., \citealt{Knapen2000}; \citealt{Fresneau2005}; \citealt{Spekkens2007}; \citealt{Castillo-Morales2007}; \citealt{Shetty2007}; \citealt{Tamburro2008}; \citealt{Garcia-Burillo2009}; \citealt{Sellwood2010}; \citealt{Meidt2013}; \citealt{Font2014b}).}

In Sect. \ref{section6}, we explained how we computed a non-circular motions map, the residual velocity field. These residual velocity fields depend significantly on the derived rotation, that is, on the rotation curve derived from the observed velocity field. Consequently, a rigorous analysis of the output of the {\sc rotcur} task in {\sc gipsy} has been carried out. We want to study both non-circular motions caused by the spiral density waves (hereafter spiral-induced non-circular motions) and those caused by the gravitational potential of the bar (bar-induced non-circular motions). The galaxies in our sample span a wide variety of morphological types and features, so a statistical approach to the non-circular motions is possible, though not straightforward.

\subsubsection{Bar-induced non-circular motions}

We are interested in analysing kinematically the effects of secular evolution by investigating the effects of the presence of a bar on the rotation of the galaxies. Our kinematic data allow us to study the kinematics and the star formation at the same time, with the caveat that we are only observing the star-forming regions. Our first goal is to understand the connection between the presence of a bar and the kinematics within the bar region. Our second goal is to study the link to star formation within the bar. 

Bars are usually dominated by old population stars, and in some cases do not show H$ \alpha $ emission (e.g., in the SB galaxies NGC~4639, NGC~5334 or NGC~5921). From the 12 SAB and {7} SB galaxies in our sample, the bar region is clearly defined in H$ \alpha $ in 11 of them. In most of these (NGC~864, NGC~1073, NGC~2500, NGC~2748, NGC~2805, NGC~4151, NGC~4389, NGC~4498, NGC~5112, NGC~6207 and NGC~6412) the bar shows H$ \alpha $ emission, and there are non-circular motions along the bar (see Table \ref{ncmtable}). Although the bar is not obvious in H$ \alpha $, using the method shown in Fig. \ref{cdf} we can in fact measure non-circular motions in three further galaxies: NGC~428, NGC~918 and NGC~3504 (Table \ref{ncmtable}). In the other galaxies (NGC~2543, NGC~2712, NGC~5740 and NGC~5921) the bar itself does not appear in H$ \alpha $, but the surrounding regions (mostly the SAR) show large non-circular motions. We have also measured the non-circular motions in the SAR in all other barred galaxies, and along the spiral arms of all galaxies. NGC~4324 is a lenticular galaxy, so there we have measured the residual velocities along the ring instead of the arms, finding values of 16.2 $ \pm $ 3.2 km s$ ^{-1} $.


{The PV diagrams along the minor axis presented in Appendix C are the first clue to see that there are deviations from circular motion. As illustrated for NGC~864 in Paper I, in the absence of non-circular motions, the velocity profile along the minor axis should be completely flat. For most of the barred galaxies in our sample, there are deviations from the systemic velocity, presented as peaks or troughs in the velocity profile along the minor axis. These indicate that some gas along the minor axis is not following the rotational pattern of the galaxy. As we see in NGC~864 (Fig.~C5), NGC~3504 (Fig.~C6) or NGC~5678 (Fig. C7), the deviations from the systemic velocity are found beyond the extent of the bar. This implies that the non-circular motions outside the bar region (i.e., in the SAR) can be a result of the influence of both potentials (bar and spiral arms).}

\begin{figure}
\begin{center}
 \includegraphics[width=84mm]{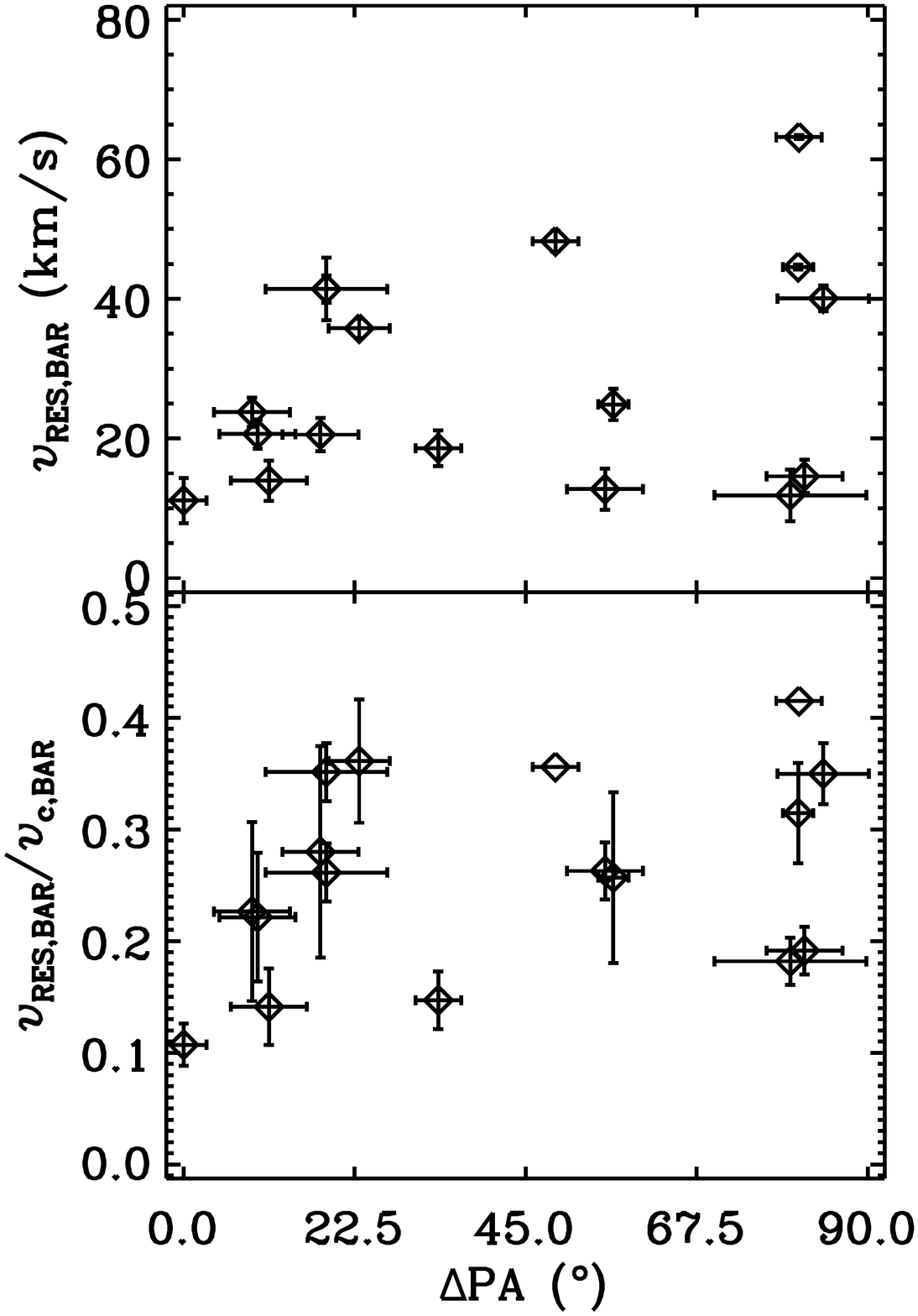}
\caption{\textit{Top)}Amplitude of the non-circular motions (residual velocites from the 95 $\%$ of the CDF) in the bar region as a function of the difference between the major axis of the bar and that of the kinematic major axis, $ \Delta $PA. Contrary to what has been reported in the literature, the highest streaming motions are not found in galaxies with $ \Delta $PA$\sim45\degr$. \textit{Bottom)} As the top panel but for the normalised residual velocities.}
\label{deltapaplots}
\end{center}
\end{figure}

{In Fig.~\ref{deltapaplots}, we represent the residual velocities in the bar region (and normalised) as a function of the difference between the major axis of the bar and that of the kinematic major axis, $ \Delta $PA. We see that the amplitude of the non-circular motions here does not increase as $ \Delta $PA approaches 45$\degr$, contrary to the expectation.}

\begin{figure*}
\begin{center}
 \includegraphics[width=168mm]{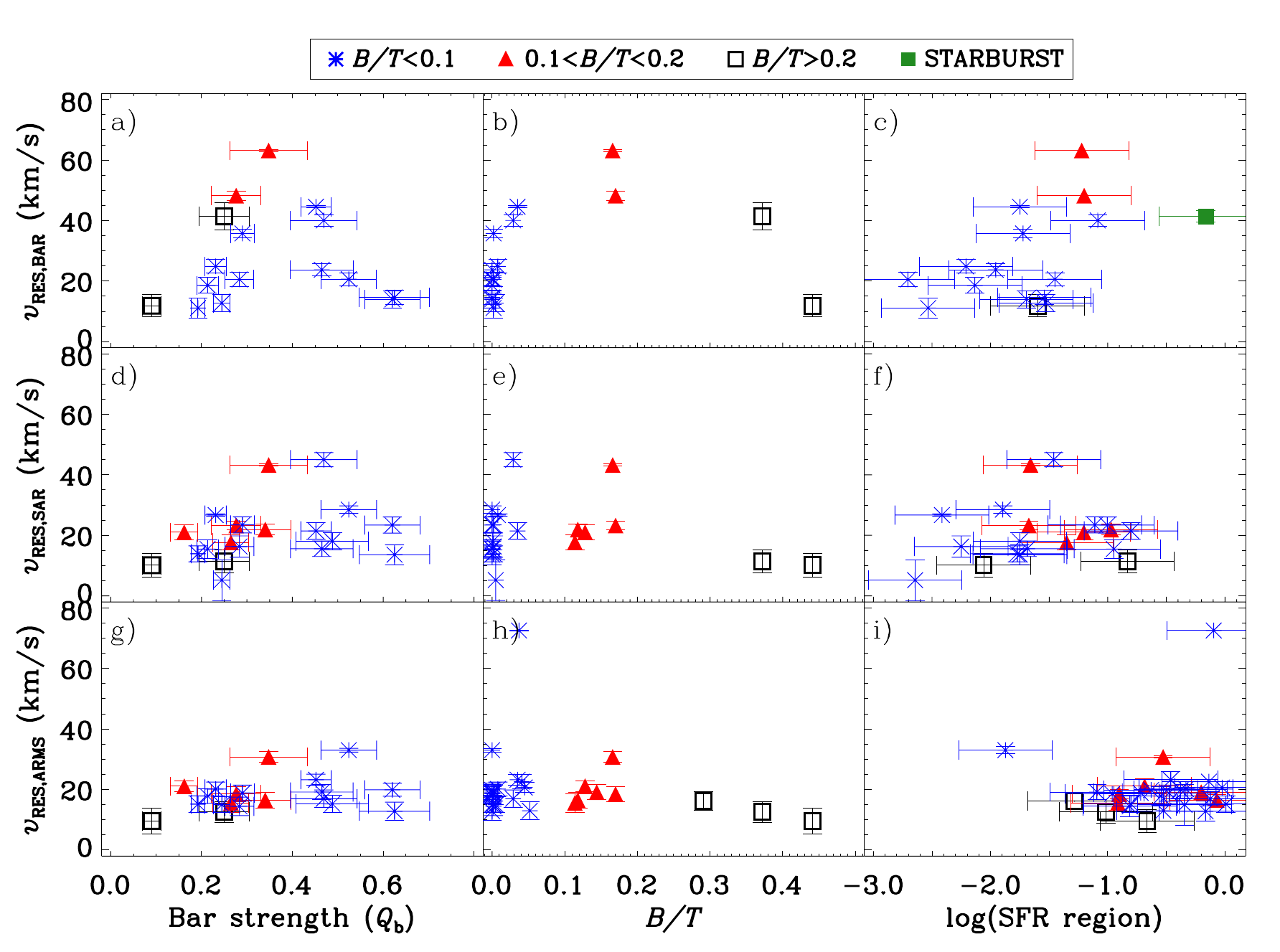}
\caption{Measurements of the non-circular motions. The top plots represent the non-circular motions (residual velocites from the 95 $\%$ of the CDF) in the bar region, whereas the middle and bottom plots represent the residual velocities in the SAR and arms respectively. \textit{Left)} Residual velocities as a function of the bar strength (represented by the parameter $ Q_{\rm b} $  from D\'iaz-Garc\'ia et al. in prep). The galaxies without a bar (i.e., $ Q_{\rm b} $=0) have not been represented. \textit{Middle)} The same residual velocities as a function of \textit{B/T}. \textit{Right)} The same residual velocities as a function of the SFR in each of the three regions. \textit{Note:} In all the plots, we have represented galaxies with \textit{B/T} $<$ 0.1 with blue asterisks, galaxies with 0.1 $<$ \textit{B/T} $<$ 0.2 with red triangles, and galaxies with \textit{B/T} $>$ 0.2 (bulge dominated) with squares. In the top right plot, we have highlighted with a filled green square the starburst nucleus (NGC~3504).}
\label{ncm1}
\end{center}
\end{figure*}

\begin{figure*}
\begin{center}
 \includegraphics[width=168mm]{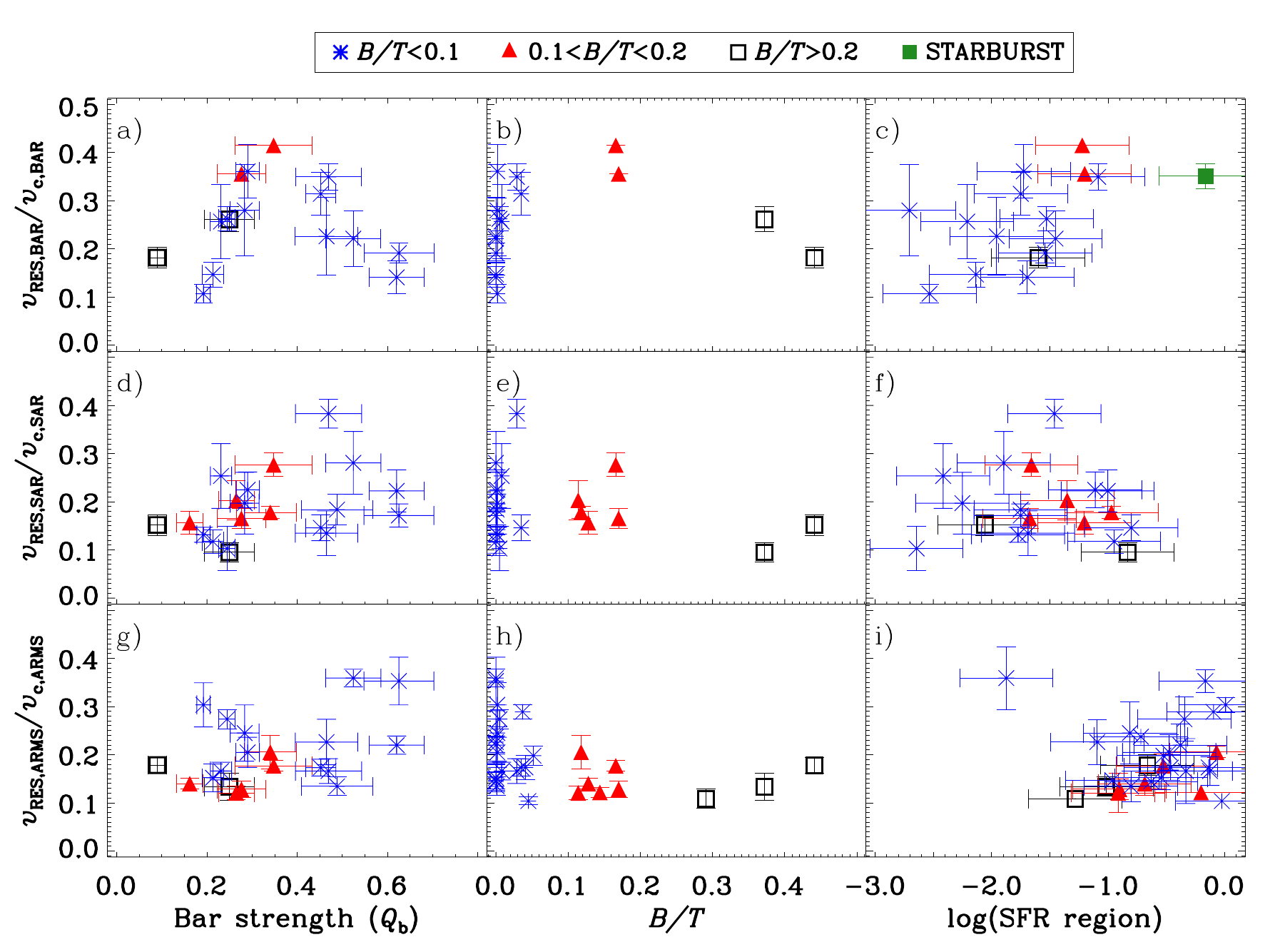}
\caption{As Fig.~\ref{ncm1} but now for the residual velocities normalised by the circular velocity at the end of the bar region (top plots), SAR (middle plots) and spiral arms (bottom plots). These circular velocities have been estimated using the universal rotation curve from \citet{Persic1991}.}
\label{ncm1bis}
\end{center}
\end{figure*}

Fig.~\ref{ncm1} (left) shows the residual velocities in the three regions -bar {(Fig.~\ref{ncm1}a)}, SAR {(Fig.~\ref{ncm1}d)} and spiral arms {(Fig.~\ref{ncm1}g)}- as a function of the bar strength, quantified by the torque parameter $ Q_{\rm b} $. $ Q_{\rm b} $ is strongly reacting to the bulge: a stronger bulge dilutes the tangential force from the bar and lowers $ Q_{\rm b} $. It is not necessarily true that the bar is weaker, just that motions are more controlled by the spherical potential of the bulge. Therefore, to distinguish the influence of the bulge in the bar strength, we have also studied the non-circular motions as a function of the bulge-to-total ratio (\textit{B/T}) in the middle plots of Fig.~\ref{ncm1}{: bar (Fig.~\ref{ncm1}b), SAR (Fig.~\ref{ncm1}e) and spiral arms (Fig.~\ref{ncm1}h)}. Also, in all the plots, we have presented the galaxies with \textit{B/T} $<$ 0.1  with blue asterisks, galaxies with $0.1 <B/T< 0.2$ with red triangles, and galaxies with $B/T> 0.2$ (bulge dominated) with squares. We list the $ Q_{\rm b} $  and \textit{B/T} values for the galaxies of our sample in Table \ref{ncmtable}. {In plots Fig.~\ref{ncm1} (a,b,c,d,e,f,g) we only represent those cases where a galaxy has a bar (i.e., $ Q_{\rm b} \neq $0).} {The same plots as in Fig.~\ref{ncm1} have been created for the normalised residual velocities, and are presented in Fig.~\ref{ncm1bis}.}

Analysing the left hand plots of Figs. \ref{ncm1} { and \ref{ncm1bis} (a,d,g)}, we conclude that $ Q_{\rm b} $ does not correlate with the amplitude of the non-circular motions. {Also, the middle plots of Figs. \ref{ncm1} and \ref{ncm1bis} (b,e,f) show no correlation between the amplitude of the non-circular motions and the presence of a bulge. Although the few cases with a significant bulge ($B/T>0.1$), show that the residual velocities in the bar decrease as \textit{B/T} increases, more data would be needed to prove whether the bulge is constraining the velocities to remain circular.}

We have measured the luminosities from the three regions identified previously in the galaxies of the sample (bar, SAR and spiral arms), and derived the SFRs in those regions using the equations presented in Sect. \ref{section7}. In panels Fig.~\ref{ncm1}{c and Fig.~\ref{ncm1bis}c}, we see a slight tendency that the higher non-circular motions correspond to star forming bars, or to bars that show more H$ \alpha $ emission (note that the scale in SFR is logarithmic). This is surprising as in the literature, the high shear and shocks in the bar region prevent star formation in the models (e.g., \citealt{Athanassoula1992b}) and observations (e.g., \citealt{Zurita2004} and \citealt{Castillo-Morales2007}), where star formation is inhibited in the bar region and the non-circular motions anti-correlate with the H$ \alpha $ luminosity there. This would cause a bias in our study, as we might not see in H$ \alpha $ those bars with higher non-circular motions. Analysing also panels {f and i of Figs.  Fig.~\ref{ncm1} and \ref{ncm1bis},} we see that the SFR and the amplitude of the non-circular motions do not correlate. 

We previously stated that some galaxies show H$ \alpha $ emission along their bar, and others do not. Why does that happen? In the literature, many studies regarding the star formation along bars have been carried out, but with no clear answer.\textit{ (i)} As we stated before, \textit{the strong shear and parallel shocks} limit the formation of stars (e.g., \citealt{Athanassoula1992b}; \citealt{Reynaud1998}; \citealt{Zurita2004}), because the giant molecular clouds can be pulled apart (\citealt{Downes1996}; \citealt{Schinnerer2002}). On the other hand, H{\sc ii} regions have been found under conditions of high shear stress and shocks (e.g., \citealt{Martin1997}; \citealt{Sheth2002}; \citealt{Zurita2008}). \citet{Martinet1997} found that the highest SFRs along the bar correspond to the stronger and longer bars in their sample of isolated late-type barred galaxies. 

\begin{figure}
\begin{center}
 \includegraphics[width=84mm]{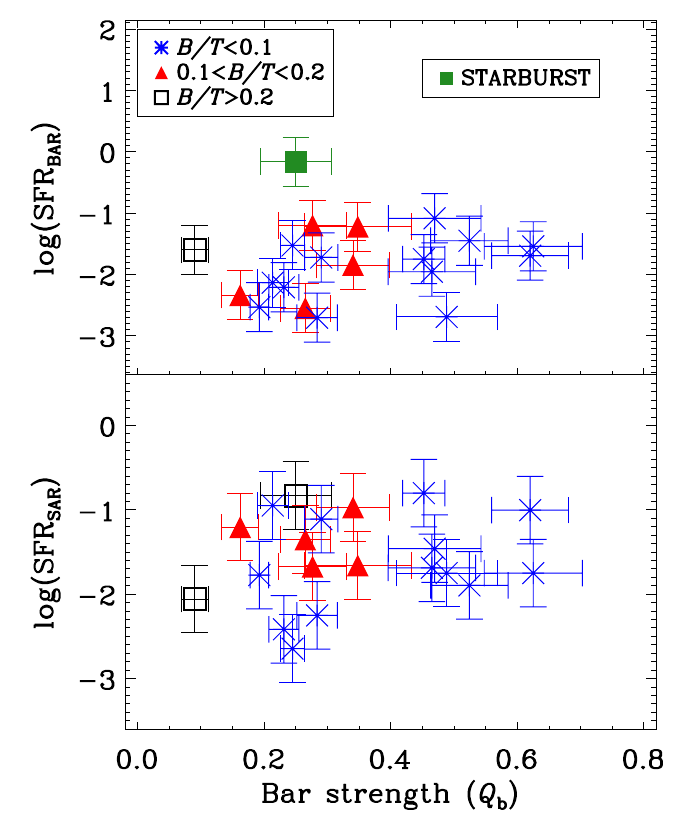}
\caption{SFRs measured in the bar region (top) and in the SAR as a function of the bar strength, represented by the parameter $ Q_{\rm b} $. We see that in this case, the strength of the bar does not regulate the H$ \alpha $ luminosity. \textit{Note:} Again, we have represented galaxies with \textit{B/T} $<$ 0.1 with blue asterisks, galaxies with 0.1 $<$ \textit{B/T} $<$ 0.2 with red triangles, and galaxies with \textit{B/T} $>$ 0.2 with squares. In the top plot, we have highlighted with a filled green square the starburst nucleus (NGC~3504) which  affects only the bar region plot.}
\label{sfrvsqb}
\end{center}
\end{figure}

In Fig.~\ref{sfrvsqb}, we present the resulting SFR values within the bar and SAR of our sample galaxies as a function of $ Q_{\rm b} $, a parameter which quantifies the strength of the bar and consequently that of shocks/shear. From Fig.~\ref{sfrvsqb} we conclude that the \textit{bar strength} does not determine the presence or quantity of H$ \alpha $ emission along the bar. In our sample, we see that the strongest bars ($ Q_{\rm b,NGC1073} $=0.63 and $ Q_{\rm b,NGC5112} $=0.62) show H$ \alpha $ emission, but further study is needed to see if this emission is located in the regions of high shocks/shear. \textit{(ii)} \citet{Garcia-Barreto1996} found that the barred galaxies with Hubble types SBa or earlier in their sample did not show H$ \alpha $ emission, probably due to the low gas content available for inflow (which is not the case, for example, for NGC~4151) so the \textit{morphological type} is not in principle the only reason. \textit{(iii)} \citet{Martin1997} also found that the two highest SFRs correspond to the galaxies which present the strongest signs of \textit{recent interaction or merging} (NGC~4731 and NGC~7479). \textit{(iv)} \citet{Sheth2000} came to the conclusion that the stars in NGC~5383 may have been formed in the spurs before the gas encounters the dust lane, and travel ballistically through the shock at the dust lane, ionizing the regions located at the leading side of the dust lane. \textit{(v)} \citet{Verley2007} suggested that the presence of star formation in the central region and along bars is a consequence of an \textit{evolutive sequence}, that goes from galaxies with H$ \alpha $ in the bar to galaxies with H$ \alpha $ emission in the circumnuclear region but not within the bar, ending in galaxies without any emission either along the bar or in a central knot.

We conclude that there is no simple answer to the question \textit{why some bars have current star formation and others don't}. The answer is probably a combination of the previous features. With our sample of galaxies, we find that neither the Hubble type nor the bar strength determine the presence of H$ \alpha $ emission within the bar region, {but larger samples are needed to establish well-based statistical results on the SF within bars. Furthermore, deeper and higher-resolution kinematic data (e.g., from the Atacama Large Millimeter/submillimeter Array, ALMA) would enable the study of the physical conditions in the bar related to the formation of stars at the scales of the molecular clouds. Then, it may be possible to understand} if a recent interaction or merger is causing the star formation of the bar, if the galaxy has already used up the available gas to be transformed into stars, or if the location of the shocks and shear correspond to regions of diminished star formation.

\subsubsection{Spiral-induced non-circular motions}
\label{spiralncm}

\begin{figure}
\begin{center}
 \includegraphics[width=84mm]{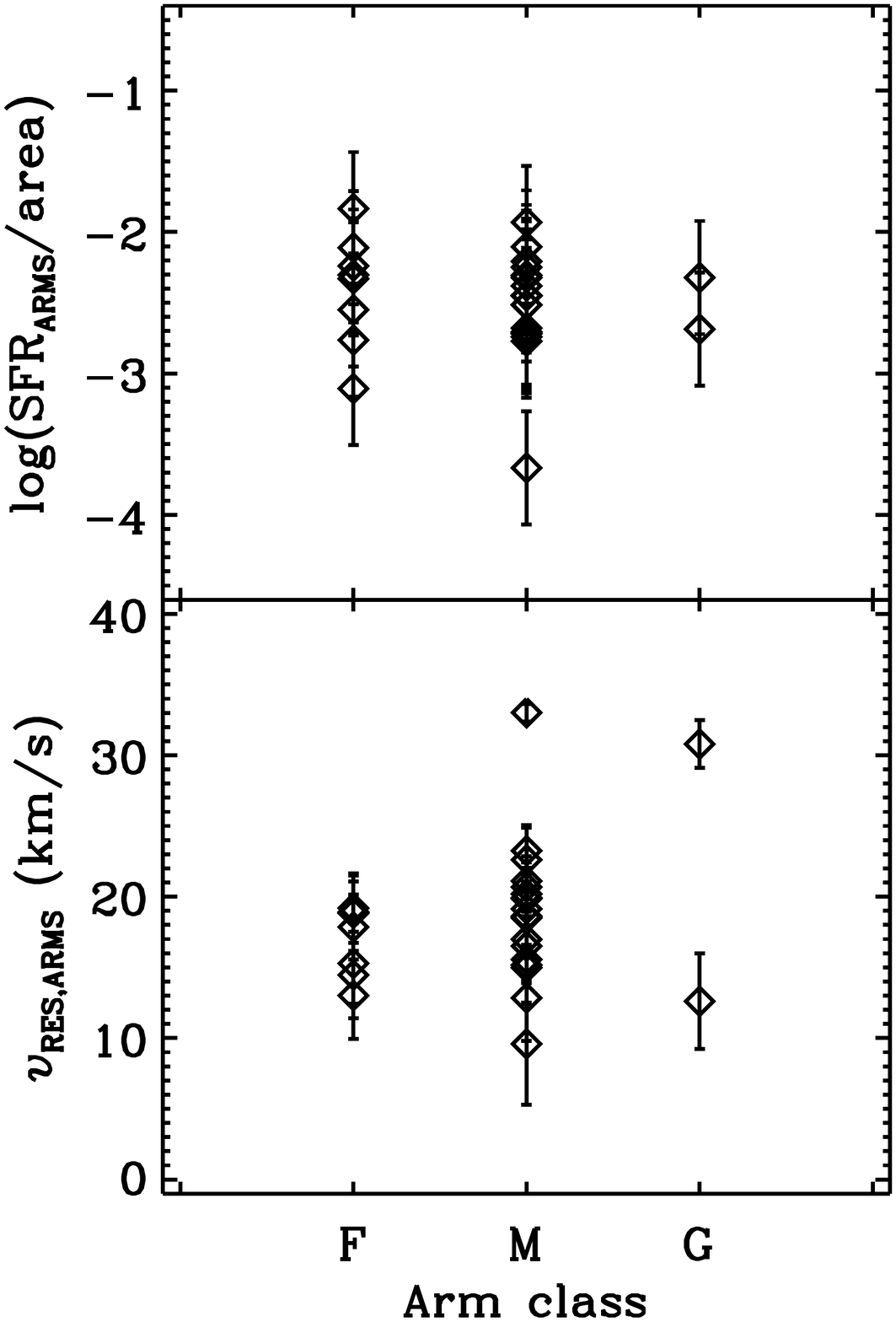}
\caption{\textit{Top)} SFR in the arms per unit area as a function of the arm class classification from {the CVRHS}. \textit{Bottom)} Amplitude of the non-circular motions within the spiral arms as a function of the arm class. We see that neither the SFR nor the amplitude of the non-circular motions correlate with the arm class. \textit{Note)}  NGC~4324 has not been represented in the figure as it does not have an arm class classification. { NGC~5678 has not been represented either because the amplitude of the non-circular motion within its spiral arms presented in Table \ref{ncmtable} is understood as being caused by the bar (see Sect. \ref{spiralncm}).}}
\label{ncm2}
\end{center}
\end{figure}

In the bar region, the mechanisms triggering star formation may be different from those in spiral arms, due to the different dynamics and shock conditions. In Fig.~\ref{ncm2}, we show star formation within the spiral arms and the amplitude of the non-circular motions there as a function of the arm class, a parameter which might plausibly be related to the ``strength" of the spiral arms. We do not find any trend or correlation. \citet{Elmegreen1986} found that the SFR per unit area does not depend on the arm class, and we confirm this.  {Therefore, we agree with other studies which confirm that the SFR does not depend on the strength of the arms (\citealt{Dobbs2009}; \citealt{Foyle2010}), and disagree with those that confirm the opposite (e.g., \citealt{Seigar2002}; \citealt{Clarke2006}).} The presence and magnitude of streaming motions in the arms seem to be a local phenomenon, unrelated to the SFR or arm class.

The high amplitude of the non-circular motions found in NGC~5678 ({with a representative value of $ \sim $70} km s$ ^{-1} $ above the average 20-30 km s$ ^{-1} $ for spiral-induced non-circular motions) leads us to consider that a spiral cannot be the single cause, and that this galaxy could be barred. This galaxy was classified as SAB in \citet{RC3} and argued to be barred by \citet{Ganda2007}, but { the CVRHS} reclassified it as SA from an S$ ^{4} $G mid-IR image and therefore does not have a value for $ Q_{\rm b}$. The PV diagram along the minor axis (Fig.~C7) shows the deviations from the circular motions, very similar to those created by the potential of the bar in NGC~864 (Paper 1). We conclude that the non-circular motions in NGC~5678 are an evidence supporting the claim that NGC~5678 is a barred galaxy, and the high amplitude of those non-circular motions could be bar-induced rather than only spiral-induced. {Another interpretation could be that there was a recent minor merger. The galaxy is asymmetric in the integrated H$ \alpha $ map, contrary to the expectations of a symmetric bar.}

\section{Conclusions}
 \label{section10}

In this paper we present the data from the kinematical study of S$ ^{4} $G galaxies which started in Paper I, and reach the following conclusions:

\begin{enumerate}
\item We have completed the observations of our kinematical study of 29 S$ ^{4} $G spiral galaxies of all morphological types using the GH$ \alpha $FaS FP instrument (FOV of 3.4 $\times$ 3.4 arcmin). The data have seeing limited angular resolution (typical values between 0.6 and 1.4 arcsec) sampled with 0\farcs2 per pixel and a spectral {sampling} $\sim$8 km s$ ^{-1} $. These FP data have been observed together with H$ \alpha $ flux-calibrated images.
\item To reach our scientific objectives, we have followed specific data reduction and analysis procedures that guarantee high angular resolution ($ \sim $1") kinematic cubes and data products.
\item The images, data described {and rotation curves} in this paper are publicly released {through the NED and the Centre de Donn\'ees Stellaires (CDS)}.
\item We flux-calibrate the GH$ \alpha $FaS data cubes following the procedures presented in Paper I. We conclude that the flux calibration can not be automated, and no standard calibration factors can be extracted for any particular GH$ \alpha $FaS filter. Instead, flux calibration is specific to each galaxy and needs to be performed by comparison with calibrated H$ \alpha $ images, such as our ACAM images.
\item We have applied the tilted-ring method to our velocity maps to extract the rotation curves. Some caveats arising in the nature of these FP data need to be taken into account (e.g., intrinsic patchiness of the line emission).
\item We have created non-circular motion maps for all the galaxies of the sample. We confirm the presence of these non-circular motions created by the non-axisymmetric potential of the bar along the bar region and at the start of the spiral arms, with a tendency that the more star-forming bars induce higher non-circular motions. We find that $ Q_{\rm b} $ does not correlate with the amplitude of the non-circular motions. However, our data is biased towards bars with recent star formation, where strong shocks and shear may take place and star formation may be inhibited. 
\item Also, we confirm the presence of non-circular motions created along the spiral arms, but there is no correlation of the amplitude of these non-circular motions with the arm class, a parameter that is related to the arm strength.
\end{enumerate}

\section*{Acknowledgments}
We acknowledge financial support to the DAGAL network from the People Programme (Marie Curie Actions) of the European Union's Seventh Framework Programme FP7/2007-2013/ under REA grant agreement number PITN-GA-2011-289313, and from the Spanish MINECO under grant number AYA2013-41243-P. This work was co-funded under the Marie Curie Actions of the European Commission (FP7-COFUND). We also gratefully acknowledge support from NASA JPL/Spitzer grant RSA 1374189 provided for the S$ ^{4} $G project. E.A. and A.B. thank the CNES for support. KS, JCMM, and TK acknowledge support from The National Radio Astronomy Observatory, which is a facility of the National Science Foundation operated under cooperative agreement by Associated Universities, Inc. This research has been supported by the MINECO under the grant AYA2007-67625-CO2-O2, and is based on observations made with the WHT operated on the island of La Palma by the Isaac Newton Group of Telescopes, in the Spanish Observatorio del Roque de Los Muchachos of the Instituto de Astrof\'isica de Canarias. The authors thank the entire S$ ^{4} $G team for their efforts in this project. We acknowledge the usage of the HyperLeda database (http://leda.univ-lyon1.fr) This research has made use of the NASA/IPAC Extragalactic Database (NED) which is operated by JPL, Caltech, under contract with NASA.

\bibliographystyle{mn2e}
\bibliography{/IAC/Papers/paper5/references}


\bsp
\clearpage


\label{lastpage}

\end{document}